\documentclass[12pt]{article}

\usepackage{graphicx}
\usepackage{slashed}
\usepackage{amsmath,amssymb}
\usepackage{bm}
\usepackage{epsfig}

\newcommand{\bpsi}{\bar{\psi}}

\begin{document}

\title{
\bf 
Light vector/axial mesons  mixings and vector meson dominance
 in  a  weak external magnetic field 
}

 \author{ F\'abio L. Braghin 
\\
{\normalsize $^1$ Instituto de F\'\i sica, Federal University of Goias, 
Av. Esperan\c ca, s/n,
 74690-900, Goi\^ania, GO, Brazil}
}

\maketitle


\begin{abstract}
Weak  magnetic field  induced 
 corrections  
to effective coupling constants 
describing  light vector mesons mixings
and 
 vector meson dominance (VMD)
are derived.
The magnetic field must be weak with respect to an effective quark mass $M^*$
such that:
$eB_0/{M^*}^2< 1$ or $eB_0/{M^*}^2<<1$.
For that, a flavor SU(2) 
 quark-quark interaction
due to  non perturbative one gluon exchange is considered.
By means of  methods usually applied to the Nambu Jona Lasinio (NJL)  and 
Global Color Models (GCM), leading
 light  vector/axial mesons couplings
 to 
a  background
 electromagnetic field are derived.
The corresponding
effective coupling constants 
are resolved in the  structureless  mesons and longwavelength   limits.
Some of the  resulting  coupling constants are
redefined such as to become
magnetic field induced  corrections 
to vector or axial  mesons couplings.
Due to the approximated chiral symmetry of the model, 
 light axial mesons mixings 
induced by the magnetic field are also obtained.
Some numerical estimates are presented 
for the 
coupling constants  and for  some of the corresponding
 momentum dependent vertices.
 The contributions of the induced VMD and   vector mesons  mixing  couplings
for the low momentum pion
electromagnetic form factor and  for the (off shell) charge symmetry violation
potential at the constituent quark level are estimated.
The  relative  overall weak magnetic field-induced   anisotropic corrections
are of the order of 
$(eB_0/{M^*}^2)^n$, where  $n=2$ or $n=1$ respectively.
\end{abstract}

%
%
%
%
%


\section{Introduction}

Light vector mesons mixings \cite{mix-1}
and vector meson dominance (VMD)
\cite{vmd,nambu-vmd}  are interesting effects considered 
 in hadron and nuclear strong interacting
systems  which are believed to
play relevant role in different processes. 
Vector mesons $\rho-\omega$ mixing
is usually attributed   to isospin violation from different quark or nucleon masses.
It 
 shows up for example in
  the pion form factor 
\cite{PPNP-thomas,thomas-pi0,dally-etal,amendolia,barkov}
and it can be responsible, at least in part, for 
  the charge symmetry violation (CSV)
component of the nuclear potential
 \cite{CSV-NN-1,CSV-NN-2,mix-np-1,mix-np-2,krein-etal-PLB,exp-rho-omega}.
An experimental value of mixing strength, 
that  is usually associated to  the energy scale of the 
rho or omega mass \cite{thomas-pi0,exp-rho-omega},
 is given by $< \rho^0 | H | \omega > = - 4520 \pm 600$ MeV$^2$.
In  the VMD  assumption
 a photon $A_\mu$  fluctuates into a quark-antiquark pair 
with the quantum numbers of a neutral rho $V_\mu^3$  or omega 
$V_\mu$ meson 
at intermediary energies. 
Different scenarios that aim to describe vector mesons dynamics 
and the electromagnetic couplings of light hadrons
were envisaged to incorporate VMD 
 \cite{PPNP-thomas,schildknecht,birse,vmd-hls,ER-1986,shakin-etal}.
From a dynamical point of view, 
 these two effects
 can be accounted by
 effective phenomenological   Lagrangian terms.
{Being} $\omega$, 
$\vec\rho$ mesons and the electromagnetic field 
 respectively denoted by $V_\mu$,  $V_\mu^i$
and $A_\mu$, the 
corresponding simplest effective couplings 
{
considered in phenomenological investigations}
can be written as
\begin{eqnarray} \label{types-couplings}
{\cal L} &=& 
g_{mix1}  {\cal F}^{\mu\nu}_3  {\cal F}^{\mu\nu} 
+
g_{mix2} V_\mu V_3^\mu 
\\
&+& g_{vmd1} F_{\mu\nu} {\cal F}^{\mu\nu}_3 
+ g_{vmd2} A_\mu V^\mu_3 
+  g_{vmd1}^\omega F_{\mu\nu} {\cal F}^{\mu\nu} 
+ g_{vmd2}^\omega A_\mu V^\mu,
\nonumber
\end{eqnarray}
where $ F^{\mu\nu},
{\cal F}^{\mu\nu}_i$ and  $ {\cal F}^{\mu\nu}$ are the 
Abelian strength  tensors for the photon, rho and omega fields respectively.
{ There are therefore  two types  of phenomenological mixing
($g_{mix1},g_{mix2}$) 
 and VMD couplings 
($g_{vmd1}, g_{vmd2}$ for the $\rho$ and 
$g_{vmd1}^\omega, g_{vmd2}^\omega$ for the $\omega$):}
 the momentum dependent ones and  the 
momentum independent ones. 
They  are usually 
 supposed to be considered
separatedly.
{
The  VMD couplings  $g_{vmd2}$ and  $g_{vmd2}^\omega$  
break gauge invariance. 
They induce  a nonzero photon mass
and therefore  they  should be avoided. 
However, 
$g_{vmd2}$ and $g_{vmd2}^\omega$
 have  been shown  to be  equivalent to
$g_{vmd1}$ and $g_{vmd1}^\omega$
  by  field redefinitions \cite{kroll-etal}.
Concerning the two different mixing couplings, several works
adopting different frameworks and methods  have shown
 the   momentum dependent coupling
turns out  to be more appropriated
 \cite{thomas-momentum-dep,CSV-NN-1,CSV-NN-2,k-dep-1,hatsuda-etal}.
}
Although the $\rho$ and $\omega$ mesons were found to be
the main vector excitations  in the chiral  SU(2) meson sector of hadrons
there are axial mesons often associated to chiral partners.
{ For instance, the   $A_1(1260)$  and the $f_1(1285)$
are usually considered  as 
chiral partners respectively  of the $\rho$   and
of the $\omega$  \cite{PDG,eLSM}. 
}
 So far
no atempt to investigate  eventual mixings  of such axial mesons has been done.

Many different effects in  hadron
structure and reactions
have been found to 
emerge in the presence of    intense  magnetic fields
expected to appear  in non central heavy ions collisions and 
  magnetars  \cite{review-B1,review-B2,review-B3,tuchin}.
These might be  relatively weak  magnetic fields with respect to 
hadron mass scales 
of the order of  $eB_0 \simeq 0.5 m_\pi^2 \simeq 0.1 {M^*}^2
\simeq 10^{17}$G, for  
 $M^* \simeq 0.33$  GeV.
Although in  peripheral heavy ions collisions $B_0$ is expected to last 
a short time 
interval  \cite{tuchin},
 vector mesons dynamics might  also be expected to have relevant effects
in these reactions.
Among the  effects produced by magnetic fields 
vector mesons  rho-omega mixing  has already  been  estimated
 \cite{mandal-etal,rho-omega-resonance-theory}.
The light vector mesons mixing,   and  also their
chiral axial mesons partners mixings,
  due to 
a weak external  magnetic field
  is addressed in the present work from a dynamical  approach.
The method considered below   also  provides $B_0$ dependent 
corrections to the VMD phenomenological  coupling  simultaneously.

In this work   
 a quark-quark interaction due to 
non perturbative one  gluon exchange,
as one of 
the leading terms for the QCD effective action, is considered.
 The one loop background field method  is applied 
and light quark-antiquark vector and axial  
mesons are introduced by means of auxiliary fields for which 
the structureless mesons limit is taken.
This approach  was able  to produce the complete
Weinberg's Large Nc  effective field theory (EFT) \cite{weinberg-2010}
for 
pions and constituent quarks, with leading and next leading symmetry
breaking terms, and their  couplings to 
background photons 
\cite{EPJA-2016,EPJA-2018}.
 Besides that,
vector/axial  mesons couplings to constituent quarks
and  their couplings to the photon have also been derived
 \cite{PRD-2018a,PRD-2018b}.
The starting Global Color Model (GCM)  is given by
the  following normalized
generating functional 
\cite{PRC1,ERV}: 
\begin{eqnarray} \label{Seff}  
Z = {\cal  N} \int  {\cal D}[\bpsi, \psi]
e^{ i \int_x  \left[
\bar{\psi} \left( i \slashed{D} 
- m \right) \psi 
-
 \frac{g^2}{2}\int_y j_{\mu}^b (x) 
{\tilde{R}}^{\mu \nu}_{bc}  (x-y) j_{\nu}^{c} (y) + \bpsi J + J^* \psi
\right] }
\;
,
\end{eqnarray}
where ${\cal N}$ is a normalization constant,
 $\int_x$ 
stands for 
$\int d^4 x$  and
$a,b...=1,...(N_c^2-1)$ stands 
for color in the adjoint representation, $N_c=3$.
{$g^2$ is the  quark-gluon coupling constant squared,}
and the color  quark currents are  given by
$j^{\mu}_a = \bar{\psi} \lambda_a \gamma^{\mu} \psi$,
being $\lambda_a$ the Pauli matrices for SU(2) isospin.
The sums in color, isospin  and Dirac indices are implicit.
The covariant quark derivative 
includes 
the minimal 
 coupling  to an external background electromagnetic field:
$
\slashed{D} = \gamma^\mu(\partial_\mu \delta_{ij} - i e Q_{ij} A_{\mu})$.
with the diagonal matrix 
$\hat{Q} =  diag(2/3, -1/3)$.
Explicit multiquark interactions in the QCD effective action
 due to gluon self interactions \cite{wang-etal}
are therefore neglected  being outside the scope of this work.
However to account for non Abelian structure of the gluon sector, the gluon propagator,
$\tilde{R}^{\mu\nu}_{ab}(x-y)$,
and the quark-gluon coupling constant will be required to be non perturbative
such that they are expected to  provide strength enough to produce
DChSB.
In several gauges 
the gluon kernel
 can be written in terms of a
transversal and a longitudinal components,
 $R_T(x-y)$ and $ R_L (x-y)$, for momentum operators in 
coordinate space as:
\begin{eqnarray}
\tilde{R}^{\mu\nu}_{ab} (x-y) 
 = \delta_{ab} \left[
 \left( g^{\mu\nu} - \frac{\partial^\mu \partial^\nu}{\partial^2}
\right)   R_T  (x-y)
+ \frac{\partial^\mu \partial^\nu}{\partial^2} R_L (x-y) \right].
\end{eqnarray}
The   method employed below was developed at length 
in Refs. \cite{EPJA-2016,PRD-2018a,PRD-2018b,EPJA-2018,PRD2016,pi-Q-B}, so that 
it will not be discussed with details  in this work.
{ 
The method is nearly equivalent to solving Schwinger Dyson equations at the 
rainbow ladder approximation with the difference here of  not considering 
the running quark effective mass from gap equation. 
The constant effective mass calculation is reliable for very low energies
and it  allows for a direct large quark mass expansion of the quark determinant.
}
In  the following section
the method will be 
schematically described
and the  sea quark determinant
will be presented for the limit of structureless vector mesons.
The determinant will be 
 expanded in the absence of  constituent quarks
in the limit
of  large quark effective mass  and small electromagnetic background field.
The leading couplings between vector/axial  mesons and photons
 will be obtained and the corresponding 
effective coupling constants will be resolved in the longwavelength limit.
{
The leading electromagnetic couplings are responsible for
weak magnetic field induced  corrections to the vector/axial mesons couplings
and the corresponding $B_0$-dependent coupling constants will be defined.
}
These coupling constants
 will be expressed in terms of the parameters of the
GCM,  Eq. (\ref{Seff}), 
and components of the quark propagator. 
However 
the momentum independent VMD terms  disappear by redefining the 
vector mesons and photon fields as shown in section (\ref{sec:vmd-Mphot})
in the same way proposed in \cite{kroll-etal}.
{ The resulting effective model are shown to be  U(1)  gauge invariant .}
{Up and down quark
masses  will be considered to be  equal} along the work to emphasize  the different nature of 
the mechanism investigated in the present work. 
{
The nature of the mixing mechanism presented below
 is not the one obtained from a type of double VMD mechanism as considered in the 
past, i.e. $\rho^0 \leftrightarrow \gamma \leftrightarrow \omega$ \cite{gasser-leutwyler}.
However it does assume
 the vector mesons to fluctuate into a  pair of quark-antiquark
  that  resolves down to a different vector meson
due to an additional electromagnetic coupling.
}
In section (\ref{sec:sub-weak}) the magnetic field dependent couplings are defined.
{ The same development is done to derive the leading pion-vector meson
couplings whose magnetic field correction will  be also tested in the pion form factor latter.}
Numerical estimates for  effective coupling constants
 and some of  the corresponding  momentum dependent form factors
 will be  shown  in section (\ref{sec:numerics}) as well as 
some simple  fittings.
Two further  applications are presented. 
Firstly the
 contributions of $B_0$ dependent VMD and 
induced mixing for the pion 
electromagnetic form factor is exhibitted in section (\ref{sec:sub-pionff}).
{
For this a phenomenological parameterization that includes VMD and vector mesons
 mixing  will be considered as discussed in Ref.  \cite{PPNP-thomas}. 
}
 And secondly,  the corresponding weak  $B_0$ induced mixing contribution 
to the (off-shell)  charge violation potential at the constituent quark level
is presented  in section (\ref{sec:sub-cvpot}).
An overal discussion will be
 presented in the final section.

\section{ Effective couplings  for light vector/axial  mesons and photons}
\label{sec:two-Q}

The main steps of the whole calculations are briefly described  in the following.
The flavor structure of the model
can be suitably investigated by means of a flavor SU(2)  Fierz transformation 
and by picking up the 
leading color singlet interaction terms as it is usually done. The 
color non singlet are smaller at least  by a factor $1/N_c$ 
besides the fact that they only generate higher order corrections
that are  numerically 
smaller
\cite{EPJA-2018}.
The quark field is splitted 
 into sea quark ($\psi_2$) 
that form light quark-antiquark states, mesons and the chiral condensate,
 and  background 
quark ($\psi_1$) 
that gives rise to
constituent quarks  eventually to compose  baryons.
At the one loop Background Field Method (BFM) level 
this  can  be done
by splitting the quark bilinears. For 
a particular Dirac  flavor-color operators $\Gamma$ 
the following shift is enough \cite{weinberg-book,EPJA-2016,EPJA-2018}:
\begin{eqnarray}
\bpsi \Gamma \psi \to  (\bpsi \Gamma \psi)_1 + (\bpsi \Gamma \psi)_2 .
\end{eqnarray}
The sea quarks can be integrated completely by means of the auxiliary field method.
Besides that, this work concerns only the vector and axial light quark-antiquark mesons 
and therefore only the corresponding vector and axial quark currents must be taken into account.
 The scalar quark interactions, however,
 is also important since it is responsible
 for the emergence of the quark-antiquark condensate 
and DChSB with the large contribution for the 
quark (hadrons) effective mass. 
The details for the scalar and pseudoscalar sector have been discussed in a large variety of 
papers and will be omitted below.

The  following
 set of bilocal  auxiliary fields (a.f.)  
is considered:
$V_\mu^i, V_\mu$, $\bar{A}_\mu^i,
\bar{A}_\mu$,
for  vectors
and axial-vectors, isospin singlets and triplets, 
{ being that the indices
$i,j,k=0,...(N_f^2-1)$, with $N_f=2$, will be used.}
They
 correspond to the color singlet vector channels 
of the  bilocal quark currents.
The following unit Gaussian  integrals for bilocal auxiliary fields are  introduced in the
generating functional \cite{PRC1,ERV,kleinert}:
\begin{eqnarray} \label{AF1}
  1 &=& N 
 \int 
D[V_\mu^i , \bar{A}_\mu^i, V_\mu, \bar{A}_\mu]
 e^{- \frac{i \alpha }{4 }
\int_{x,y} 
(\bar{R}^{\mu\nu})^{-1} ( 
 V^i_{\mu} V^i_{\nu}
+ 
   \bar{A}^i_{\mu}  \bar{A}^i_{\nu} )
}
 e^{- \frac{i\alpha}{4 }  
\int_{x,y}   (\bar{R}^{\mu\nu})^{-1} 
(
 V_{\mu} V_\nu
+ 
\bar{A}_{\mu}   \bar{A}_{\nu} 
)
},
\end{eqnarray}
 where  $N$ is a normalization constant that does not show up in observables and
\begin{eqnarray}
\bar{R}^{\mu\nu} \equiv \bar{R}^{\mu\nu}(x-y)= g^{\mu\nu} ( R_T(x-y) + R_L(x-y) )
+ 2 \frac{\partial^\mu \partial^\nu}{\partial^2 } ( R_T(x-y) - R_L(x-y) ) ,
\end{eqnarray}
and $\alpha=4/9$. 
{
Now  it is possible to introduce the renormalization constants and 
to perform
a shift in each of the auxiliary field  with the corresponding
 quark-current with the same quantum number
 given by:
\begin{eqnarray}
V^i_{\mu} \to
 (Z_V^{\frac{1}{2}} V^i_{\mu} - 
  g  Z_g Z_\psi     {\bar{R}_{\mu\nu}}  {j^{V,\nu}}^{i,(2)} ) ,
\;\;\;\;\;\;
 \bar{A}^i_{\mu} \to 
 ( Z_A^{\frac{1}{2}}
\bar{A}^i_{\mu} -  g  Z_g Z_\psi    {\bar{R}_{\mu\nu}}  {j^{A,\nu}}^{i,(2)} ) ,
\nonumber
\\
V_{\mu}  \to (
Z_V^{\frac{1}{2}} V_{\mu} -  g  Z_g Z_\psi    {\bar{R}_{\mu\nu}}   {j^{V,\nu}}^{(2)}), 
\;\;\;\;\;\;\;
   \bar{A}_{\mu} 
\to 
  ( Z_A^{\frac{1}{2}}
 \bar{A}_{\mu} -  g  Z_g Z_\psi   {\bar{R}_{\mu\nu}}  {j^{A,\nu}}^{(2)}  ) ,
\end{eqnarray}
where the usual quark   flavor currents
 were considered \cite{PRD-2018a}:
$j_{V,\mu}^{i,(2)} = \bpsi \gamma_\mu \lambda^i \psi$,
$j_{V,\mu}^{(2)} = \bpsi \gamma_\mu  \psi$,
$j_{A,\mu}^{i,(2)} = \bpsi \gamma_5 \gamma_\mu \lambda^i \psi$
and 
$j_{A,\mu}^{(2)} = \bpsi \gamma_5 \gamma_\mu  \psi$.
These shifts have clearly unity Jacobian.
{The wavefunction renormalization constants
will not however  be written explicitely along the 
development below  until section (\ref{sec:vmd-Mphot}).}
The (sea) quark vector/axial quark-current  interactions from the 
Fierz transformation
are canceled out  and it becomes possible to integrate out sea quarks.

However for the low energy regime these bilocal fields can be
expanded in an infinite basis of local meson  fields \cite{PRC1}.
 By picking up only the lowest energy modes 
and making the form factors  
to reduce to constants in the zero momentum  limit 
it yields the structureless mesons limit analysed in Refs.
\cite{PRD-2018a,PRD-2018b}.
{
The vector mesons and their chiral partners were assumed to 
develop the same normalization constants that provide the 
corresponding  canonical normalization.
}
With this, the components of the gluon propagator in Eq. (\ref{AF1})
are absorved in the structureless vector/axial mesons normalization constants.
The saddle point equations for the auxiliary fields  are  the usual gap equations
and only the scalar one might have non trivial solution corresponding to
the quark-antiquark scalar condensate from
DChSB 
 in the vacuum, $< S >$.
 By taking into account  this constant,
the  free quark  propagator  with a suitable implicit regularization, 
by omitting the quark wavefunction
renormalization constant,
can then be written as:
 $$
S_{0,c} (x-y) = \left( i \slashed{D} -  M^* 
\right)^{-1}  \delta (x-y)
$$ 
where
 $M^* = m +   <S>$.
Its behavior in the vacuum and finite energy density systems has been investigated 
in many works.
In particular, the chiral condensate and consequently  the quark effective mass $M^*$
have strong dependence on an external  magnetic field 
\cite{review-B1,review-B2,review-B3}.
 This $B_0$ dependence will be investigated in  the numerical analysis  done below 
by choosing two different values for the effective mass $M^*$, one 
in the vacuum ($M^* = 330$  MeV) and another larger 
value for finite (weak) $B_0$.

The  Gaussian integration of the sea   quark field is performed and
the resulting determinant can be written
as:
\begin{eqnarray} \label{Seff-det}  
S_{eff}   &=&   i \; Tr  \; \ln \; \left\{
- i \left( S_{c}^{-1} (x-y) 
+
\sum_q  a_q \Gamma_q j_q (x,y) \right)
 \right\} 
,
\end{eqnarray}
where 
$Tr$ stands for traces of all discrete internal indices 
and integration of  spacetime coordinates
and  the
 quark kernel can be written as 
\begin{eqnarray}
S_{c}^{-1} (x-y) =  
S_{0,c}^{-1} (x-y) + \Xi_v  (x-y),
\end{eqnarray}
where the following quantity with the (chiral)
structureless vector/axial mesons fields 
was defined above:
\begin{eqnarray} \label{Xi-q}
\Xi_v (x,y)    =
-   \frac{\gamma^\mu}{2} \left[   
 \sigma_i
 \left(     V_{\mu}^i (x)
+   \gamma_5 
 \bar{A}_{\mu}^i (x) \right)
+    
 \left(     V_{\mu} (x)
+   \gamma_5 
 \bar{A}_{\mu} (x) \right) 
  \right] \delta(x-y),
\end{eqnarray}
with 
canonically normalized fields and where
{
$\sigma_i$ are  the isospin Pauli matrices}.
All the constituent quark currents were included in 
following quantity $\sum_q  a_q \Gamma_q  j_q  (x,y)$, 
for the
 particular (Dirac, flavor) $q$-channel.
 Their couplings to the light vector mesons and to the pseudoscalar/scalar sector
were investigated  respectively in Refs.\cite{PRD-2018a,PRD-2018b}
and \cite{EPJA-2016,EPJA-2018,PRD-2019,pi-Q-B} 
 and they  will be neglected in this work.

\subsection{
Leading  
photon-vector mesons  from expansion}

Consider a large quark effective mass of the determinant above 
{ within a zero order derivative expansion  \cite{mosel},
for the  longwavelength
 regime such that the local limit is reached.}
{
To understand how the couplings and coupling constants were resolved, consider
the following term  involving  
vector meson and electromagnetic fields
from  the first order of the expansion:
\begin{eqnarray}
I_{2nd-a} &=&
i 2 d_1  \; Tr \; \left[ 
S_{0,c} (x-y) \gamma^\mu \sigma_i V_\mu^i (y)
\; 
S_{0,c} (y-z) \gamma^\rho  e \hat{Q} A_\rho (z) \right],
\end{eqnarray}
where 
$d_n =   \frac{(-1)^{n+1} }{2 n}$.
Within the zero order derivative expansion, the traces in color, isospin and Dirac
are taken and 
this yields the coupling constant $g_{\rho A}$ below.
}
 The following  leading and next leading local  terms are obtained:
\begin{eqnarray} \label{I-VMD}      
{\cal L}_{VMD} &=& -
g_{\rho A}  V^\mu_3 A_\mu
-
g_{\omega A}  V_\mu A^\mu
-
g_{F\rho} {\cal F}_{\mu\nu}^3 F^{\mu\nu}
- 
g_{F\omega} {\cal F}_{\mu\nu} F^{\mu\nu} ,
\\
 \label{FFF-1}   
 {\cal L}_{F}
&=&
g_{F\rho \omega}
(  F^{\mu\nu} {\cal F}_{\nu\rho}^3 {\cal F}^\rho_\mu
+  F^{\mu\nu} {\cal G}_{\nu\rho}^3 {\cal G}^\rho_\mu
)
-
g_{FF\omega} 
F_{\mu\nu} F^{\nu\rho}
{\cal F}_\rho^{\mu} 
- g_{FF\rho} 
F_{\mu\nu} F^{\nu\rho}
{\cal F}_\rho^{3, \mu} 
\nonumber
\\
 &+&
\epsilon_{ij3} \left[ 
g_{m1} 
F_{\mu\nu} V_i^\mu V_j^\nu
+
g_{m1A} 
 F_{\mu\nu} \bar{A}_i^\mu \bar{A}_j^\nu
\right]
-
 \epsilon_{ij3}
\left[ 
 g_{m1}
{\cal F}^i_{\mu\nu}
 A^\mu V^\nu_j
+ g_{m1A} 
{\cal G}_{\mu\nu}^i
 {A}^\mu \bar{A}_j^\nu
\right]
,
\end{eqnarray}
where  the following Abelian strength tensors have been defined:
\begin{eqnarray} \label{tensors-1}
{\cal F}^{\mu\nu} = 
\partial^\mu V^\nu
-  \partial^\nu V^\mu,
\;\;\;\;\;\;\;
{\cal F}^{\mu\nu}_i =
\partial^\mu V^\nu_i
-  \partial^\nu V^\mu_i,
\\
 \label{Gmunu-i}
{\cal G}^{\mu\nu} = \partial^\mu \bar{A}^\nu
-  \partial^\nu \bar{A}^\mu
\;\;\;\;\;\;\; 
{\cal G}^{\mu\nu}_i 
= \partial^\mu \bar{A}^\nu_i
-  \partial^\nu \bar{A}^\mu_i.
\end{eqnarray}
Although the present approach gives rise to non Abelian corrections
for the mesons field strength  
\cite{meissner}
these interaction terms are outside of the scope of the present work.
The effective coupling constants in (\ref{I-VMD},\ref{FFF-1}) 
were resolved in the long wavelength limit  
by calculating the traces in Dirac, color  and isospin indices.
After a Wick rotation to the Euclidean momentum space,
they can be written as:
\begin{eqnarray} \label{g-rho-B}
 g_{\rho A} &=& 3 g_{\omega A}
 =   4 \;  e \;  N_c \;  d_1  \; Tr' \; 
((  \tilde{S}_2 (k)   )),
\\ \label{g-F-rho-B}    
 g_{F\rho} = 3 g_{F \omega} &=& 
 8 \; e \; N_c  \;  d_1    \; Tr' \; 
((\tilde{S}_0  (k) \tilde{S}_0 (k)   )),
\\
\label{FFFvmd}  
g_{FF\rho} = \frac{3}{5} g_{FF\omega}
&=& \frac{4 e}{3} g_{F\rho\omega}
 =
\frac{8}{3} \; e^2 \; N_c \;  d_1      
  \; Tr' \; 
(( \tilde{S}_0 (k)  \tilde{S}_0 (k) \tilde{S}_0 (k) ))
,
\\ \label{mix-A}    
g_{m1}  &=&
 4 \; e \; N_c \;  d_1      
  \; Tr' \; 
(( \tilde{S}_0 (k) \tilde{S}_2 (k) ))
,
\\
\label{mix-AAb}
g_{m1A} &=&
g_{m1}
+ {M^*}^2  d_1 8 N_c \; Tr' \; 
((  \tilde{S}_0 (k)  \tilde{S}_0 (k) \tilde{S}_0  (k)))
= 
g_{m1}
+ 4 {M^*}^2  g_{F\rho \omega}
,
\end{eqnarray}
where 
 $Tr'(( ..))$ are integrals  in internal momenta for the zero momentum exchange limit.
The following functions  were  used:
\begin{eqnarray}   \label{S0S2}
\tilde{S}_0 (k) =
\frac{1}{ k^2 + {M^*}^2 },
\;\;\;\;\;\;\;\;
\tilde{S}_2 (k) = \frac{k^2 - {M^*}^2}{ (k^2 + {M^*}^2)^2 }.
\end{eqnarray}
The interactions  in Eq. (\ref{I-VMD})
 correspond to the two different 
types of  vector meson dominance coupling.
 The first coupling constants $g_{\rho A},g_{\omega A}$,
with dimension $M^2$ where $M$ is a mass scale,
 are ultraviolet (UV) divergent and they
 are related  by:
$f_r \equiv  g_{\omega A}/g_{\rho A} = 1/3$.
{ This ratio
is close to usual fittings  $1/f_r \sim  3.5$ }\cite{thomas-momentum-dep}.
{ Note that
 the definition obtained  in the present work is the inverse of the 
usual convention for these couplings,  i.e. $g_{\rho A} \sim 1/g_{\rho}$.}
The same ratio holds for the momentum dependent coupling constants
$g_{F\omega}/g_{F\rho} = 1/3$ being that these couplings are 
 dimensionless and logarightmic 
UV divergent.
The way to handle  UV divergences  will be discussed below.
The  dimensionless  coupling constants
$g_{m1}$ and $g_{m1A}$  are  also logarithmic divergent and they  correspond to
  electromagnetic couplings of vector and axial mesons.

{
The two first couplings, $g_{\rho A}$ and 
$g_{\omega A}$,
 that break U(1)  gauge invariance will be  eliminated 
by field redefinition in Section (\ref{sec:vmd-Mphot}).
}
{
 Note
however the two types of couplings $g_{m1}$ and $g_{m1A}$ written separatedly in the 
Eq. (\ref{FFF-1}): the first involves the electromagnetic tensor whereas the second do not.
The first ones are U(1) gauge invariant whereas the second are not.
However it is interesting to note that 
these last couplings $g_{m1}$ can be made gauge invariant 
by considering further terms of the expansion 
and writing   the original covariant derivative of the 
quark propagator
$$
-
 \epsilon_{ij3}
\left[ 
 g_{m1}
{\cal F}^i_{\mu\nu}
 \tilde{D}_\mu V^\nu_j
+ g_{m1A} 
{\cal G}_{\mu\nu}^i
 \tilde{D}^\mu \bar{A}_j^\nu
\right],
$$
where  $\tilde{D}_\mu = \partial_\mu + e_v A_\mu$ was defined
by extracting a factor $e_v$ outside from the definition of $g_{m1}$.
In this covariant derivative,
the actions of the charge operator on quark and vector mesons 
appear as a trace ($tr_I$) in isospin indices $ e_v = e \;  tr_I (\sigma_i Q \sigma_j)$.
This  corresponds to an averaged electrical charge of the 
overall  effective coupling.
This more complete parameterization
 will not be considered below since it does not yield leading contributions
for the VMD and mixing form factors as desired.
}
The coupling constants $g_{F\rho\omega}$,
$g_{FF\rho}$ and $g_{FF\omega}$  have  dimension $M^{-2}$
and they  can be seen as mixings
induced by an external photon
and effective 
couplings of each of the vector mesons
 to two photons.
{These issues will be reminded in the conclusions and 
they  might have interesting consequences.}
These last coupling constants are UV finite and 
proportional  among each other
and they will provide the most important results of this work.

In Figures (1-a) and (1-b)
the Feynman diagrams corresponding to the couplings
 (\ref{g-rho-B},\ref{g-F-rho-B}) are shown.
The photon is represented by a  wavy line, the vector
mesons by a dotted-dashed line,
and internal quarks  by solid lines.
The couplings for the strength tensors of vector mesons or photon  are indicated by 
a triangle in the corresponding vertex.
In Figures  (2-a),   (2-b) and (2-c)
  the Feynman diagrams for all the interaction terms
(\ref{FFF-1})
are exhibited.
{
 In Fig.2 it is seen that although the  mixing involves 
a neutral vector meson (say $\rho^0$) fluctuating into a pair of quark-antiquark,
an electromangnetic  coupling is needed for the 
fluctuating quark-antiquark to annihilate  into a different
vector meson (say $\omega$).

}

\begin{figure}[ht!]
\centering
\includegraphics[width=80mm]{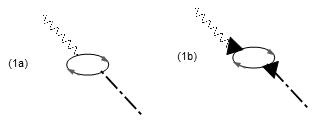}
\caption{ \label{fig:diagrams1}
\small
In these diagrams,
an internal  solid line represents a sea   quark, 
the wavy and dot-dashed lines stand for a photon and a vector meson respectively.
The wavy (dot-dashed) line with a triangle in a vertex stands for the 
electromagnetic (vector mesons) strength tensor $F^{\mu\nu}$
(${\cal F}^{\mu\nu}_i, {\cal F}^{\mu\nu}$ ).
}
\end{figure}

\begin{figure}[ht!]
\centering
\includegraphics[width=100mm]{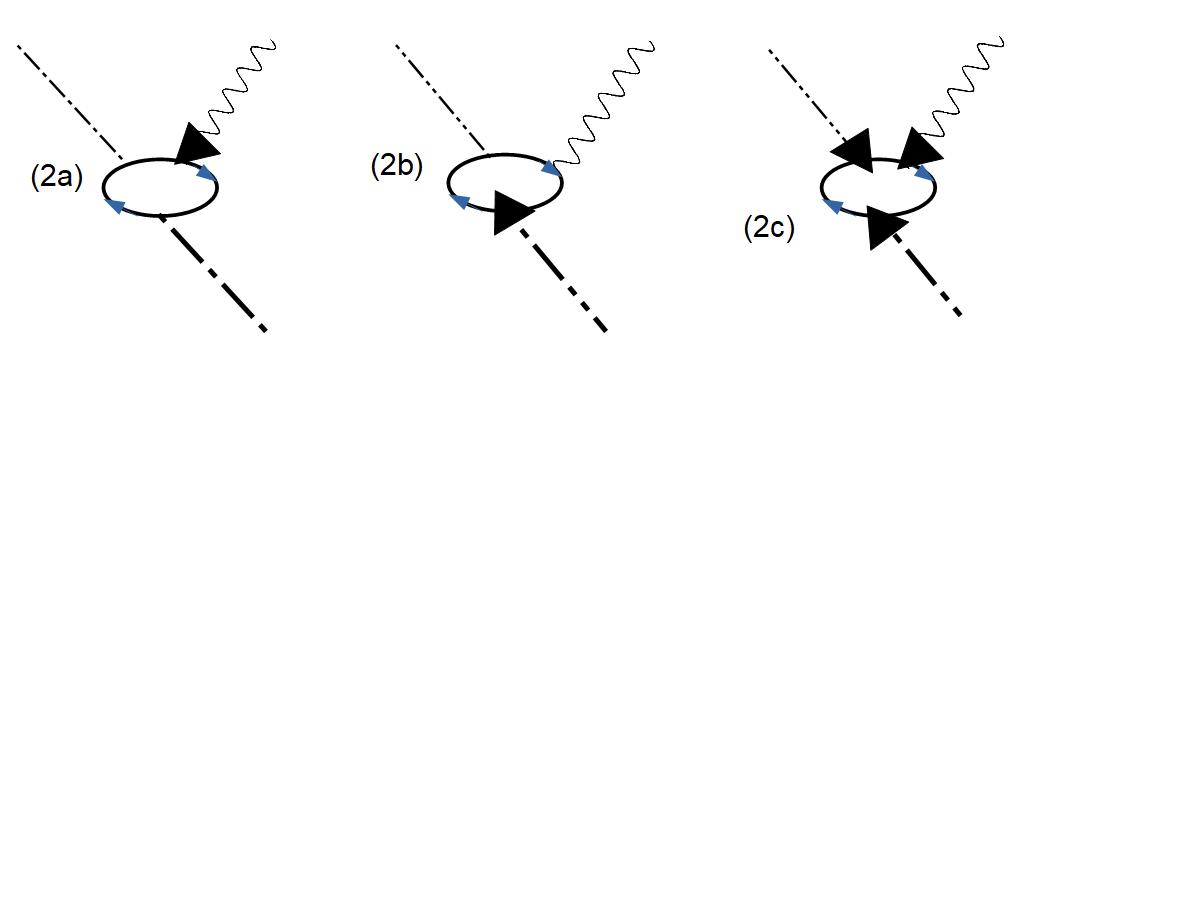}
\caption{ \label{fig:diagrams2}
\small
In these diagrams,
the same convention used in Figure (\ref{fig:diagrams1})
is considered for the three-leg vertices.
}
\end{figure}

\subsection{ Some leading pion-vector mesons couplings}

{
Although this work  is not dedicated to calculating  pion couplings,
there is one relevant pion coupling that will be considered
in the pion form factor   below 
as test for the magnetic field dependent
 VMD and vector mesons mixings couplings.
By considering the pion field  from pseudoscalar auxiliary fields, 
in
the usual  suitable non linear realization,  
its contribution in determinant (\ref{Seff-det}) can be written by
exchanging
the function $\Xi_v$ by $\Xi_v + \Xi_s$ where
$$
\Xi_s = F \left( U (x) + U^\dagger (x) \right),
$$
where
$F$ is the pion normalization constant,
identified to the $F_\pi$,
 $U (x) = e^{i \vec\pi (x) \cdot \vec\sigma/F}$ being $\vec{\pi} (x)$ the  local pion field
\cite{EPJA-2016,EPJA-2018,PRD-2019}.
Within the same large quark effective  mass expansion the following 
leading 
gauge (U(1)) invariant couplings to vector mesons and  
electromagnetic  field  appear. 
Without calculating all  the traces in isospin indices the leading terms can be
written as:
\begin{eqnarray} \label{rhopi1a}
 \Delta L_{\rho \pi}^{(1)} &=&
\bar{g}_{\rho\pi1} \; tr_I \left[ \left(\partial_{\mu}  - i e \hat{Q} A_\mu \right) 
\sigma_i \sigma_j \sigma_k \right]
(\pi_i \pi_j) V_k^\mu
+ \bar{g}_{\omega\pi1} 
 \; tr_I \left[ \left(\partial_{\mu}  - i e \hat{Q} A_\mu \right) 
\sigma_i \sigma_j  \right]
 \pi_i \pi_j  V^\mu
\nonumber
\\
 &+& 
{g}_{F\rho\pi1} T_{ijk} 
 F_{\mu\nu} 
(\partial^\mu  \pi_i \pi_j) V_k^\nu
+
{g}_{F\omega\pi1} \; tr_I  (\hat{Q} \sigma_i \sigma_j ) F_{\mu\nu} (\partial^\mu {\pi_i \pi_j} )
 V^\nu
,
\end{eqnarray}
 where  
the isospin tensor is
 $T_{ijk} = \delta_{i3} \delta_{jk} - \delta_{j3} \delta_{ik} + \delta_{k3} \delta_{ij}
+ \frac{i}{3} \epsilon_{ijk}$.
From these terms, a covariant derivative for the pion and vector mesons can be defined
as  $\bar{D}_\mu = \partial_\mu - \bar{e} A_\mu$
where  $(\bar{e} \; T_{ijk}) =
tr_I (\hat{Q} \sigma_i \sigma_j \sigma_k)$  for the first term (and the same for the 
third term) includes   an averaged charge.
These different components might contribute in each of  the channel of 
the vertex that  is considered and this tensor structure   garantees
electric  charge conservation for the particular channel in consideration.
In particular the neutral-rho and pion coupling in the pion form factor below corresponds
to the term $\delta_{k3} \delta_{ij}$.
Because the higher order  derivative couplings are  not of interest in the following part of the work,
the pion-vector meson and electromagnetic field  coupling will be redefined
to incorporate $\bar{e}$,
for example:  $g_{\rho\pi1} = \bar{e}_\rho \bar{g}_{\rho\pi1}$
and  $g_{\omega\pi1} = \bar{e}_\omega \bar{g}_{\omega\pi1}$,
being $\bar{e}_\rho$ and $\bar{e}_\omega$ are
 not the rho and  omega charge but an averaged value
in the corresponding vertex.
Therefore  we can write for the electromagnetic couplings from Eq. (\ref{rhopi1a}):
\begin{eqnarray} \label{rhopi1}
 \Delta L_{\rho \pi} &=&
{g}_{\rho\pi1} \;  T_{ijk} A_\mu 
{\pi}_i \pi_j   V_k^\mu
+ {g}_{\omega\pi1} 
 \;   A_\mu 
 \vec\pi^2  V^\mu
\nonumber
\\
 &+& 
{g}_{F\rho\pi1} T_{ijk} 
 F_{\mu\nu} 
(\partial^\mu  \pi_i \pi_j) V_k^\nu
+
{g}_{F\omega\pi1}  F_{\mu\nu} (\partial^\mu {\vec\pi^2 }) V^\nu
,
\end{eqnarray}
where in Euclidean momentum space:
\begin{eqnarray}
g_{\rho\pi1}  = 3 g_{\omega\pi1} =  \frac{M^*}{F} g_{m1} ,
\;\;\;\;\;
 g_{F\rho\pi1} =  3 g_{F\omega\pi1} =  \frac{M^*}{F} g_{F\rho\omega} .
\end{eqnarray}
It is interesting to note that 
the coupling $\rho-\pi-A_\mu$ is  enhanced 
 with respect to the $F_{\mu\nu}-\rho-\omega$ coupling
by a constant factor $M^*/F$.
The  lowest order derivative terms, with $F_{\mu\nu}$, are leading 
and their values are directly proportional to electromagnetic  couplings to vector mesons,
 along the lines of the Universality hipothesis.
There are several other couplings of the pions which however are outside the scope of this work.

}

\subsection{ Equivalence of VMD couplings and zero photon mass}
\label{sec:vmd-Mphot}

The momentum dependent VMD couplings in Eq.  (\ref{I-VMD})
have  been found more suitable to describe
 phenomenology  \cite{thomas-momentum-dep}.
Besides that  it has been pointed out 
by Kroll et al \cite{kroll-etal}
that a field redefinition eliminates 
the momentum independent ones.
To see that, 
 consider the resulting free  vector/axial mesons terms ${\cal L}_{free}$
\cite{PRD-2018a,PRD-2018b} and the leading electromagnetic  
 correction terms ${\cal L}_{A,M}$ from the expansion of the determinant above:  
\begin{eqnarray} \label{Ifree}
{\cal L}_{free} =
- \frac{g_f^{(0)}}{4} \left( {\cal F}^{\mu\nu}_i {\cal F}_{\mu\nu}^i 
+ 
{\cal G}^{\mu\nu}_i {\cal G}_{\mu\nu}^i
+ 
{\cal F}^{\mu\nu} {\cal F}_{\mu\nu}
+
{\cal G}^{\mu\nu} {\cal G}_{\mu\nu}
\right)
,
\\
 \label{A-mass}
{\cal L}_{A,M} = -  \frac{5}{9}\frac{{M_v^{(0)}}^2}{2} 
\frac{e^2}{f_v^2} A_\mu A^\mu 
- \frac{g_{F}}{4}
F_{\mu\nu} F^{\mu\nu}
-
\frac{M_v^{(0)}}{2} \left( {V_\mu^i}^2 
+
{V_\mu}^2 \right) 
-  \frac{ {M_{\bar{A}}^{(0)} }^2 }{2}  \left( {\bar{A}_{i,\mu}}^2
+ {\bar{A}_\mu}^2
\right), 
\nonumber
\end{eqnarray}
where $f_v$ is the vector mesons
 field normalization and
 the following effective parameters have been defined in the same 
long wavelength and 
zero momentum limit considered before:
\begin{eqnarray} \label{gf-vecmes} 
g_{f}^{(0)} 
&=&
 d_1 4 N_c \; Tr' \; 
(( \tilde{S}_0^2 (k))),
\\  \label{mass-vecmes} 
{M_v^{(0)}}^2
&=&
d_1 8  N_c \; Tr' \; 
(( \tilde{S}_2 (k) )),
\\
\label{mass-axmes} 
{M_A^{(0)}}^2
&=&
{M_v^{(0)}}^2
+ 4 g_f^{(0)} {M^*}^2
,
\end{eqnarray}
By  considering    field redefinitions of the type:
\begin{eqnarray}
{A}_\mu = c \tilde{A}_\mu, \;\;\;\;
V_\mu^3 = \tilde{V}_\mu^3  + \frac{c e}{2 f_v} \tilde{A}_\mu, 
\;\;\;\; V_\mu = \tilde{V}_\mu + \frac{c e}{6 f_v} \tilde{A}_\mu,
\end{eqnarray}
where $c$ is a constant,
the photon mass reduces to zero and the momentum independent 
VMD couplings
$g_{\rho A},g_{\omega A}$ disappear from the 
Lagrangian  
in the same way it was 
 shown in Ref. \cite{kroll-etal}.

{ 

Several of the expressions found above, Eqs.
(\ref{g-rho-B},\ref{g-F-rho-B},\ref{mix-A},\ref{gf-vecmes},\ref{mass-vecmes},\ref{mass-axmes}),
are ultraviolet (UV) divergent and the following procedure was adopted to render results finite.
There are only two types of divergences, quadratic and logarithmic,
 and they show up in Eqs. 
(\ref{gf-vecmes},\ref{mass-vecmes}).
These two quantitites are simply expected to 
provide  the free vector mesons ($\rho$ or $\omega$)
tree level parameters, mass and field  normalization.
By fixing the values of ${M_v^{(0)}}^2, g_f^{(0)}$, 
one may expect 
to attribute finite values for the two types of integrals in all the expressions above.
All the other UV divergent coupling constants can be written in terms of these
 integrals 
and with that  they become finite.
Therefore, let us consider finite fixed values:
\begin{eqnarray}    \label{prescriptions}
g_f^{(0)} &=& f_v^2,
\nonumber
\\
{M_v^{(0)}}^2 &=& m_V^2 .
\end{eqnarray}
Initially one could take $F_v = 1$ and $M_v^{(0)}\simeq0.77$  GeV.
However smaller values should be expected  because to obtain
Eqs.  (\ref{gf-vecmes}) and (\ref{mass-vecmes})
the vector mesons structureless limit has been assumed. 
{
With this, the axial mesons masses are different from the vector mesons ones by 
a finite positive factor as shown in Eq. (\ref{mass-axmes}).
By considering the values used below ($g_f^{(0)}\simeq 0.5$ and $M^*=0.33$ GeV),
 the axial meson-vector mesons
mass difference is of the order of $\Delta M \sim \sqrt{2} \times 0.33 = 0.467$ GeV that 
provides the correct experimental mass differences values  \cite{PDG}.
}
When plugging these relations in the coupling constants with UV divergences, 
 Eqs.  (\ref{g-rho-B},\ref{g-F-rho-B},\ref{mix-A}) and (\ref{mix-AAb}),
it can be written that:
\begin{eqnarray} \label{grhoA-reg}
g_{\rho A} = 3 g_{\omega A}
 =
 \frac{e}{2} m_V^2
,
\\ \label{g-F-rho-B-reg}  
 g_{F\rho} = 3 g_{F\omega} =
2 \; e  \; f_v
,
\\ \label{mix-A-reg} 
g_{m1}  =
 e   \; f_v
\; + \;  
\frac{3  {M^*}^2}{e}  g_{FF\rho} 
= e f_v + 4 {M^*}^2 g_{F\rho \omega} ,
\\
\label{mix-A-reg-A}
g_{m1A} =
e \; f_v + \frac{6 {M^*}^2}{e} g_{FF\rho}
= e \; f_v + 8 {M^*}^2 g_{F\rho \omega}
.
\end{eqnarray}
This
 procedure might be viewed as a (partial) renormalization of the model
and the reason is the following. 
{The  renormalized quantities are obtained with implicit
$Z_q, Z_\rho, Z_\omega,Z_A$ 
(quark, rho, omega and axial meson 
wavefunction renormalizations) 
and eventually 
$Z_g$ for the 
quark-gluon coupling constant.
These renormalization constants make 
all the divergent integrals finite
in the same way.
{
Eqs. (\ref{gf-vecmes}) and (\ref{mass-vecmes}) may  be considered
renormalization conditions for the isospin symmetric case analysed in the present work
with   $Z_\rho= Z_\omega=Z_A$.
The effective coupling constants depend on the  
 relevant renormalization constants
basically in the 
same way at this order of the quark determinant expansion, so that
 it is enough to consider the two renormalization conditions above
to render all the UV-divergent  integrals finite.}
Eqs. (\ref{grhoA-reg}-\ref{mix-A-reg}) 
establish well definite scales or order of magnitude for the
resulting effective  coupling constants in terms of $g_f^{(0)}$ and 
$M_v^{(0)}$.

}

\subsection{ Weak magnetic field depedent effective couplings }
\label{sec:sub-weak}

{For 
a  magnetic field,
$B_0$   along the 
$- \hat{z}$ direction
 one might choose
$A^\mu =  B_0 (0, 0, x, 0)$.
 The three-leg coupling constants
in Eq. (\ref{FFF-1})
can be then
 rewritten to incorporate $B_0$.
 By ideintifying explicitely the Lorentz and isospin components 
 index of the 
tensors and fields, the following terms arise:}
\begin{eqnarray} \label{VMD-B}
 \label{FFF-2}
{\cal L}_{F}^B
&=&
 \left(\frac{e B_0}{{M^*}^2}\right)  g_{F\rho\omega}^B
(   {\cal F}_{\mu=2, \rho}^{i=3} {\cal F}^{\rho}_{\nu=1}
+  {\cal G}_{\mu=2, \rho}^{i=3} {\cal G}^{\rho}_{\nu=1}
)
-
 \left(\frac{e B_0}{{M^*}^2}\right)  g_{FF\rho}^B 
  F^{\mu=2,\rho}
{\cal F}_\rho^{i=3,\nu=1} 
- 
 \left(\frac{e B_0}{{M^*}^2}\right)  g_{FF\omega}^B
 F^{\mu=2,\rho}
{\cal F}_\rho^{\nu=1} 
\nonumber
\\
&+&
  \left(\frac{e B_0}{{M^*}^2}\right)  \epsilon_{ij3} 
[ - g^B_{m1}
{\cal F}^{i=3}_{\mu=2,\nu}
 V^\nu_j
- g^B_{m1A}  {\cal G}_{\mu=2,\nu}^i
  \bar{A}_j^\nu
+   
 g_{m2}^B V_i^{\mu=1} V_j^{\nu=2}
+ 
g_{m2A}^B
\bar{A}_i^{\mu=1} \bar{A}_j^{\nu=2}
]
,
\end{eqnarray}
where some of  these effective coupling constants 
were simply redefined as:
\begin{eqnarray}  \label{ff-relations} 
g_{FF\rho}^B = \frac{3}{5} g_{FF\omega}^B
= \frac{4 e}{3}  g_{F\rho\omega}^B  =
\frac{ {M^*}^2}{e} g_{FF\rho} ,
\\  \label{gm2B} 
 g_{m2}^B =
  \frac{{M^*}^2}{e}  g_{m1} ,
\\  \label{gm2B-A} 
 g_{m2A}^B = 
\frac{{M^*}^2}{e}  g_{m1A} ,
\\
 \label{mix-A-B}
 g_{m1}^B  =
  4 \; N_c \;  d_1      {M^*}^2
  \; Tr' \; 
(( \partial_{q_x^{(1)}} [\tilde{S}_0 (k) \tilde{S}_2 (k,k+q+q_1) ] ))_{q_{(1)}=0}
,
\\
\label{mix-A-B-A}
g_{m1A}^B =
g_{m1}^B 
+ 4  {M^*}  g_{F\rho\omega}^B ,
\end{eqnarray}
where $q^{(1)}$ is the momentum eventually carried by the electromagnetic field, 
 $\partial_{q_x^{(1)}} = \frac{\partial}{\partial q_x^{(1)} }$
and 
\begin{eqnarray}
\tilde{S}_2 (k,k+q)= \frac{ (k^2 + k \cdot q - {M^*}^2)}{
(k^2+{M^*}^2) ((k+q)^2+{M^*}^2)}.
\end{eqnarray}
The effective coupling constants  $g_{m1}^B$ 
and $g_{m2}^B$ have  dimensions $M$ and $M^2$ respectively,
while the others $g_{F\rho\omega}^B$ 
and $g_{FF\rho}^B, g_{FF\omega}^B$
are dimensionless.
The coupling constants $g_{F\rho\omega}^B$
 is  a magnetic field induced
anisotropic vector mesons mixing  term.
{
 It is also in  the axial mesons mixing coupling 
in Eq.
(\ref{mix-A-B-A}).
}
Note that the difference in the $\omega$ and $\rho$ vector mesons  
 couplings to the electromagnetic field  in the vacuum ($g_{F \rho}, g_{F\omega}$)
is not the same under weak magnetic field due to the isospin/chiral symmetry breakings
induced by the magnetic field.
 According to expressions above, the ratio of the anisotropic corrections  to 
VMD effective couplings 
due to weak $B_0$ is 
$$ 
\frac{g_{FF\rho}^B}{g_{FF\omega}^B}
= \frac{3}{5}.
$$
{ 
Note however  that the  
strengths of the
 axial  meson mixing coupling constants
are  larger than the corresponding
ones from the  vector  mesons 
mixings. 
}

$g_{m1}^B, g_{m1A}^B$ and 
$g_{m2}^B, g_{m2A}^B$
 correspond respectively to  
anisotropic  magnetic field  corrections 
to  the  propagation of 
charged rho and A$_1$ mesons that mix Dirac components $x,y$.
{ 
{The terms  from $g_{m_2}^B$  and $g_{m 2A}^B$ 
might be identified as types of  mass-corrections  that }
 can be written as:  
\begin{eqnarray}  \label{M-xy}
\Delta M_{\rho_{\pm}}^{x,y} = \sqrt{\frac{eB_0}{{M^*}^2}}
\sqrt{\frac{2 g_{m2} {M^*}^2}{e}},
\;\;\;\;\;\;
\Delta M_{\bar{A}_{\pm}}^{x,y} = \sqrt{\frac{eB_0}{{M^*}^2}}
\sqrt{\frac{2 g_{m2A} {M^*}^2}{e}}.
\end{eqnarray}
}

 It is interesting to emphasize that  the weak magnetic field corrections to
mixing and VMD couplings are simply proportional among themselves.
Therefore the magnetic field induced coupling constants
have  two contributions, one from the quark effective mass $M^*$ that 
is dependent on the magnetic field from the  gap equation
and  the multiplicative factor $eB_0/{M^*}^2$.
The dependence of all these quantities on the up and down 
quark mass differences will not be considered at length   here.

{
Concerning the pion couplings to electromagnetic field  and to the vector mesons in 
Eq. (\ref{rhopi1}) only  the leading terms will be considered: those whose couplings
are proportional to $F_{\mu\nu}$. 
The B-dependent correction to the leading pion-vector mesons couplings
(\ref{rhopi1})    
can then be written as:
\begin{eqnarray} \label{pion-rho-B}
L_{\rho\pi}^B &=&
g_{F\rho\pi1}^B \frac{e B_0}{{M^*}^2}  i \epsilon_{ijk}   
(\partial^x \pi_i \pi_j) V_k^{(2)}
+
g_{F\omega\pi1}^B   \frac{e B_0}{{M^*}^2}
  \partial^x \vec{\pi}^2 V^{(2)}
,
\end{eqnarray}
$$
g_{F\rho\pi1}^B =  3 g_{F\omega\pi1}^B 
 = \frac{{M^*}^2}{e}  g_{F\rho\pi1}  = \frac{{M^*}^3 }{e F} g_{F\rho\omega} =
\frac{M^*}{F}  g_{F\rho\omega}^B.
$$
}

\section{ Numerical estimates and form factors}
\label{sec:numerics}

Exact relations between some of the effective 
coupling constants were already exhibited 
in Eqs. (\ref{g-rho-B},\ref{g-F-rho-B}) and (\ref{FFFvmd})
and the corresponding 
$B_0$-dependent ones  (\ref{ff-relations},\ref{gm2B}).
 Further relations for  resulting effective parameters  and the regularization parameters,
$M_v^{(0)}, g_f^{(0)}$,
were shown in Eqs. (\ref{grhoA-reg},\ref{g-F-rho-B-reg}) and (\ref{mix-A-reg}).
 Some  simple  approximated ratios 
can also be estimated
in the limit of very large quark effective mass by noting the 
dependence of the quark kernels on $M^*$, i.e. 
$$S_0 \sim \frac{1}{M^*}, 
 \;\;\;\;\; 
\tilde{S}_0 \sim \frac{1}{{M^*}^2}.$$
This yields:
\begin{eqnarray}
\label{approx-ratios}
\frac{g_{F\rho\omega}}{g_{m1}}
 \sim 
 \frac{1}{2 {M^*}^2},
\;\;\;\;\;\;
\frac{g_{\rho A}}{g_{F \rho }} \sim 
 \frac{ {M^*}^2 }{2},
\;\;\;\;\;\;
\frac{g_{F\rho\omega}^B}{g_{F\rho } } 
\sim
\frac{B_0 }{4 {M^*}^2}.
\end{eqnarray}
These ratios
make explicit the relative order of magnitude of the effective coupling constants
in the limit of very large $M^*$.
The second of these ratios, for the VMD momentum 
dependent and independent coupling constants, 
 have the same order of magnitude of
the corresponding ratio calculated with 
values fitted from phenomenology.

In the Table I several estimations for coupling constants in the vacuum and under weak 
$B_0$ from Eqs. 
(\ref{g-rho-B},\ref{g-F-rho-B},\ref{mix-A}) and (\ref{gm2B},\ref{M-xy}) 
are  presented for different values of 
the   normalization
 integrals (\ref{prescriptions}).
The only coupling constant 
with explicit normalization dependence on the effective mass $M^*$ is
$g_{m_2}^B$, from Eq. (\ref{mix-A-reg}).
 In this case, values for two different effective mass 
were presented.
The structureless limit of the vector mesons can be expected 
to be responsible for limitted account of the  vector mesons
 mass
and normalization constant, by means of the 
parameters $M_v^{(0)}$ and $g_f^{(0)}$.
Therefore  smaller  values than  
the expected ones for a tree level Lagrangian terms to  describe vector mesons dynamics
were also considered.
{
To compare with results, usual values for the
  VMD  phenomenological  coupling constants are the following:
$$ g_{vmd 1}= \frac{e}{g_\rho} \sim 6.2 \times  10^{-2},
\;\;\;\;\;\;
 g_{vmd 2} \simeq  \frac{e m_\rho^2}{ g_{\rho}}\simeq 3.7 
 \times  10^{4} \; MeV^2. $$
}  
\cite{shakin-etal}.
The best values for the two  normalization  parameters
can be expected to be $m_V \simeq 0.5$   GeV and
$f_v \simeq 0.1$ for Eqs. (\ref{grhoA-reg}) and (\ref{g-F-rho-B-reg})
to reproduce respectively $g_{vmd 2}$ and $g_{vmd 1}$ from phenomenology.

\begin{table}[ht]
\caption{
\small
Numerical results for several of the coupling constants   in the vacuum and induced by
weak $B_0$  from Eqs.  (\ref{g-rho-B},\ref{g-F-rho-B}) and (\ref{mix-A}) 
and (\ref{gm2B},\ref{M-xy})
respectively.
Different values of the {\it normalization} values  for 
the integrals from $f_v, m_V$
given by  Eq. (\ref{prescriptions}).
} 
\centering 
\begin{tabular}{c c c c c c} 
\hline\hline 
coupling constants & \;  $f_v=1.0$, &
$f_v=0.1$,   
   &
$f_v=0.1$,   & $f_v=1.0$  & $f_v=1.0$,
\\
$m_V$ (GeV)   &   $0.77$     &  $0.77$ 
  & $0.5$     &   $0.5$
  &   $0.1$ 
\\
\hline
$g_{\rho A}$  (MeV$^{2}$) \; &  
$8.9 \times 10^{4}$  &    $8.9 \times 10^{4}$   &     $3.8 \times 10^{4}$   &    $3.8 \times 10^{4}$
  &   $1.5 \times 10^{3}$
 \\ [0.5ex]
$g_{F \rho}$ \;\;\;\;\;\;\;\;\;\;\;\;\;  
  &  0.606  &  0.061   &  0.061    &  0.606    & 0.606
\\ [0.5ex]
$g_{m_1}$    \;\;\;\;\;\;\;\;\;\;\;\;\;   
 &  0.315  & 0.042    &  0.042     &  0.315   &   0.315
 \\ [0.5ex]
$g_{m_2}^B$  (GeV$^2$)
  &  0.113  &  0.015    &  0.015    &    0.113  &0.113 
\\ [0.5ex]
 ($M^*=0.33$ GeV)  & & & & & 
 \\ [0.5ex]
$g_{m_2}^B$  (GeV$^2$)
  & 0.210  &  0.032   &  0.032     &  0.210    &  0.210
 \\ [0.5ex]
 ($M^*=0.45$  GeV) 
& & & & & 
  \\ [0.5ex]
$\frac{\Delta M_\rho^{x-y}}{\sqrt{(eB_0/{M^*}^2)}}$  (MeV) 
  & 475  & 173  & 173  &  475 & 475
 \\[1ex]
($M^*=0.33$ GeV)
& & & & & 
 \\[1ex] 
\hline 
\end{tabular}
\label{table:results} 
\end{table}

The couplings  $g_{FF\rho}^B$ and $g_{FF\omega}^B$
represent  corrections to
vector  meson dominance terms induced
 by the external magnetic  field and it contains an anisotropic contribution.
These terms can be added to the terms in (\ref{I-VMD})
for the VMD coupling constant.
Similarly the emerging magnetic field induced anisotropic 
mixing term has a coupling with $g_{F\rho\omega}^B$.
The resulting coupling constants, with anisotropic corrections,
 can be written as:
\begin{eqnarray}
g^{vmd1}_\rho (B_0) &=&
 g_{F\rho } +
 \left(\frac{e B_0}{{M^*}^2}\right) 
g_{FF\rho}^B  
,
\\
g^{vmd1}_{\omega} (B_0) &=& g_{F\omega } +
 \left(\frac{e B_0}{{M^*}^2}\right)  g_{FF\omega}^B ,
\\
g_{mix1} (B_0) &=&  \left(\frac{e B_0}{{M^*}^2}\right)  
 g_{F\rho\omega}^B
\end{eqnarray}
where the resulting coupling constants are dimensionless
and they still present the implicit magnetic field dependence of 
$M^*$.

Considering two different values for the quark effective mass 
$M^*$ the following values are 
obtained:
\begin{eqnarray} \label{gFFrho33}
M^*=0.33 \,  \mbox{GeV} &\to&  g_{FF\rho} = 1.07 \times 10^{-2} \,\mbox{GeV}^{-2},
\\
&& g_{FF\rho}^B  = 3.83 \times 10^{-3}, \;\;\;\;\; g_{F\rho \omega}^B = 2.48 g_{FF\rho}^B,
\nonumber
\\ \label{gFFrho45}
M^*=0.45 \mbox{GeV} &\to&  g_{FF\rho} = 5.74 \times 10^{-3} \, \mbox{GeV}^{-2},
\\
&& g_{FF\rho}^B  = 3.83 \times 10^{-3}, \;\;\;\;\; g_{F\rho \omega}^B = 2.48 g_{FF\rho}^B.
\nonumber
\end{eqnarray}
The weak magnetic field induced  corrections to the 
momentum dependent VMD and 
vector mesons mixings  are therefore finite and basically independent of $M^*$.
These  coupling constants, 
$g_{FF\rho}^B\propto g_{F\rho\omega}^B$,
are of the order of $(eB_0/{M^*}^2)$, i.e.  therefore
 small with respect to the zero magnetic field value
 $g_{F\rho}$.

{
After the estimates done above, some further numerical ratios between coupling constants presented
in this work will be exhibitted next.
In particular ratios between magnetic field induced corrections to rho and omega mesons dominance
and also vector mesons and axial mesons
coupling constants.
By taking numerical values for $M^* = 0.33$ GeV
and $ eB_0/{M^*}^2 = 0.1$ one has:
\begin{eqnarray}    \label{ratio-gvmd1}
R_1  \equiv  \frac{  g_{vmd1} (B_0)  }{ g_{vmd1}^\omega (B_0)   } =
\frac{ g_{F\rho} + (\frac{eB_0}{{M^*}^2})  g_{FF\rho}^B }{ 
 g_{F\omega} + (\frac{eB_0}{{M^*}^2})  g_{FF\omega}^B   }
& \sim & 2.99 \;\;  (\mbox{for}\;\; f_v=1.0)
\nonumber
\\
  &\sim&  2.93   \;\;\;\;  (\mbox{for} \;\; f_v = 0.1)
,
\\      \label{ratio-mixA-A}
R_3 \equiv \frac{  g_{m2A}^B }{ g_{m2}^B   } = \frac{g_{m1A}}{ g_{m1}}
 = \frac{e f_v + 6 \frac{{M^*}^2}{e} g_{FF\rho} }{e f_v + 3 \frac{{M^*}^2}{e} g_{FF\rho} }
&\sim&  1.04 \;\;\;\;  (\mbox{for}\;\; f_v=1), 
\nonumber
\\ 
 &\sim& 1.27 \;\; (\mbox{ for}\;\; f_v=0.1) .
\end{eqnarray}
The ratio $R_1$ can be  compared to its value in the vacuum discussed 
after Eq. (\ref{S0S2}) that is $R_1 (B_0=0) = 3$.
From ratio $R_3$
 the momentum independent axial mesons mixing coupling constant ($g_{m2A}^B$)
induced by weak  magnetic field
 is larger than the corresponding
 vector mesons mixing coupling constant  ($g_{m2}^B$).
The momentum dependent mixings $g_{F\rho\omega}^B$ were noted 
to be the same at this level of calculation in Eq. (\ref{VMD-B}).

}

\subsection{ Form factors}

{
Next, 
it is interesting to write the Fourier transformation 
of 
  complete   non local  expressions for some of the terms 
 obtained from the quark determinant expansion
 (\ref{I-VMD},\ref{FFF-1}) and also (\ref{FFF-2}).
Consider the following terms:
\begin{eqnarray} \label{ff}
{\cal L}_{ff} &=& -
\bar{g}_{\rho A} (x-y)  V^\mu_{i=3} (x)  A_\mu (y)
-
\bar{g}_{F\rho}  (x-y) {\cal F}_{\mu\nu}^{i=3} (x) F^{\mu\nu} (y)
+ \bar{g}_{FF\rho}  (x,y,z)
F_{\mu\nu} (x)  F^{\nu\rho} (y)
{\cal F}_\rho^{i=3, \mu} (z)
\nonumber
\noindent
\\
 &+&
\epsilon_{ij3}
\left[ 
 \bar{g}_{m1} (x,y,z)
F_{\mu\nu} (x)  V_i^\mu (y)  V_j^\nu (z)
+ 
 \bar{g}_{m1A} (x,y,z)
 F_{\mu\nu} (x)   \bar{A}_i^\mu (y)  \bar{A}_j^\nu (z)
\right]
\nonumber
\\
&+&
 \bar{g}_{FF\rho}^B (x-y)
\frac{eB_0}{{M^*}^2}
F^{\nu=2,\rho} (x)
{\cal F}_\rho^{\mu=1, i=3}  (y)
.
\end{eqnarray}
The resulting expressions are the
momentum dependent  vertices that  correspond to 
the Feynman diagrams of Figures (\ref{fig:diagrams1},\ref{fig:diagrams2}).
 The following  Fourier transformed terms  are obtained:
}
\begin{eqnarray} \label{ff}
\tilde{\cal L}_{ff} &=& -
g_{\rho A} (Q^2) V^\mu_{i=3} (Q) A_\mu (- Q)
-
g_{F\rho} (Q^2) {\cal F}_{\mu\nu}^{i=3} (Q) F^{\mu\nu} (-Q)
\nonumber
\\
&+& g_{FF\rho}  (Q,Q_2)
F_{\mu\nu} (Q_2) F^{\nu\rho} (Q)
{\cal F}_\rho^{i=3, \mu} (Q+Q_2)
\nonumber
\\
 &+&
\epsilon_{ij3}
\left[   
g_{m1} (Q,Q_2)
F_{\mu\nu} (Q_2)  V_i^\mu (Q) V_j^\nu (Q+Q_2)
+ 
 g_{m1A} (Q,Q_2)
 F_{\mu\nu} (Q_2)  \bar{A}_i^\mu (Q) \bar{A}_j^\nu (Q+Q_2)
\right]
\nonumber
\\
&+& g_{FF\rho}^B (Q^2) 
\frac{eB_0}{{M^*}^2}
F^{\mu=2,\rho} (Q)
{\cal F}_\rho^{\nu=1,i=3}  (-Q)
,
\end{eqnarray}
{
 In these equations $\pm Q$ is a (incoming/outgoing)
 vector or axial meson momentum and $Q_2$
is the momentum carried by the electromagnetic field whenever it carries momentum.}
The corresponding form factors,
 in  Euclidean  momentum space,
are given by: 
\begin{eqnarray} \label{ff-rho-B}
g_{\rho A}
(Q^2)  &=& 4 e N_c d_1    \; \int_k 
\frac{k^2  + k \cdot Q - {M^*}^2}{( k^2+ {M^*}^2)
( (k + Q)^2 + {M^*}^2)},
\\
\label{ff-F-rho-B}
g_{F \rho} (Q^2) &=&   8 e  N_c d_1  \; \int_k 
\frac{1}{   (k^2+ {M^*}^2)
( (k + Q)^2 + {M^*}^2) },
\\
\label{ff-m1}
g_{m1} (Q,Q_2)
&=&
4 e N_c d_1  \; \int_k
\frac{ ( k^2 + k \cdot Q + k \cdot Q_2 - {M^*}^2 ) }{
(k^2 + {M^*}^2 ) ( (k+Q_2)^2 + {M^*}^2)
(  (k+ Q + Q_2)^2  + {M^*}^2)
}
,
\\
\label{ff-m1A}
g_{m1A} (Q,Q_2)
&=&
g_{m1} (Q,Q_2) 
+ \frac{ 3 {M^*}^2 }{e}    g_{FF\rho} (Q,Q_2)
,
\\
g_{FF\rho} (Q,Q_2) 
 &=& 
\frac{8}{3} \; e^2 \; N_c \;  d_1      
 \; \int_k
\frac{1}{  (k^2+ {M^*}^2)
( (k + Q)^2 + {M^*}^2) ( (k+Q+Q_2)^2 + {M^*}^2)   }
,
\\
\label{ff-FF-B}    
g_{FF\rho}^B
(Q^2)  &=&
  8 e {M^*}^2  N_c d_1 \; \int_k 
\frac{1}{( k^2+ {M^*}^2)^2
( (k+ Q )^2  + {M^*}^2) }
,
\end{eqnarray}
where $\int_k= \int d^4 k/(2\pi)^4$.
{
The form factors were presented as functions of $Q^2$ or 
$(Q,Q_2)$ being this second case functions of all the scalars
formed by the momenta $Q$ and $Q_2$.}
{ The coupling constants were define in the 
 zero momentum exchange limit 
 of these expressions  ($Q=Q_2=0$). }
 Although the integrals of Eqs. (\ref{ff-rho-B},\ref{ff-F-rho-B}) and (\ref{ff-m1})
have  quadratic or 
 logarithmic UV divergences, 
their values at $Q=0$
 are normalized 
 according to the prescriptions (\ref{prescriptions})
 as discussed above.
All the curves shown below for the form factors do not 
include the linear momentum $Q$ from the strength tensors
in Eq. (\ref{ff}).

 In  Figure (\ref{fig:3}) the following  form factors 
  $g_{\rho A}(Q^2), g_{F\rho}(Q^2)$ 
and also 
$g_{FF\rho }^B(Q^2)$
 are drawn
for $M^*= 330$ MeV.
The coupling   $g_{FF\rho}^B(Q^2)$
 represents the weak magnetic field correction 
to VMD and it is proportional to the induced vector mesons mixing strength,
{ from Eqs. (\ref{FFF-1},\ref{FFFvmd})}:
$$g_{F\rho\omega}^B(Q^2) = \frac{3}{4e} g_{FF\rho}^B (Q^2).$$
{
 As a consequence there is an overal linear dependence 
of the coupling   $g_{F\rho\omega}^B$ on the 
magnetic field, besides the effective mass $M^*$ dependence on $B_0$.}

Simple approximated fittings for the three form factors of figure (\ref{fig:3}), $g_{\rho A}(Q^2)$,
 $g_{F\rho}(Q^2)$ and $g_{FF\rho}^B(Q^2)$, were found 
with the following simple functions parameterized by two constants
each of them ($k_{\rho A}, k_{F\rho}, k_{FF\rho}$
and $g_{\rho A}(0), g_{F\rho}(0), g_{FF\rho}^B (0)$):
\begin{eqnarray}
Fit_1 (Q^2)  &=& 
k_{\rho A}^4 \frac{g_{\rho A}(0)}{(Q^2 + k_{\rho A}^2)^2},
\\
Fit_2  (Q^2) &=&
 k_{F\rho }^2 \frac{g_{F\rho }(0)}{(Q^2 + k_{F \rho }^2)},
\\
 Fit_3(Q^2) &=& 
 k_{FF\rho}^4 \frac{g_{FF\rho}^B(0)}{(Q^2+k_{FF\rho}^2)^2}.
\end{eqnarray}
The zero momentum values in
  these fittings are those from the Table and (\ref{gFFrho33})
and approximated values for the other constants are the following:
$k_{\rho A} \simeq 2.2$ GeV, 
$k_{F\rho } \simeq 1.8$ GeV 
and $ k_{FF\rho} \simeq  1.5$ GeV.

\begin{figure}[ht!]
\centering
\includegraphics[width=140mm]{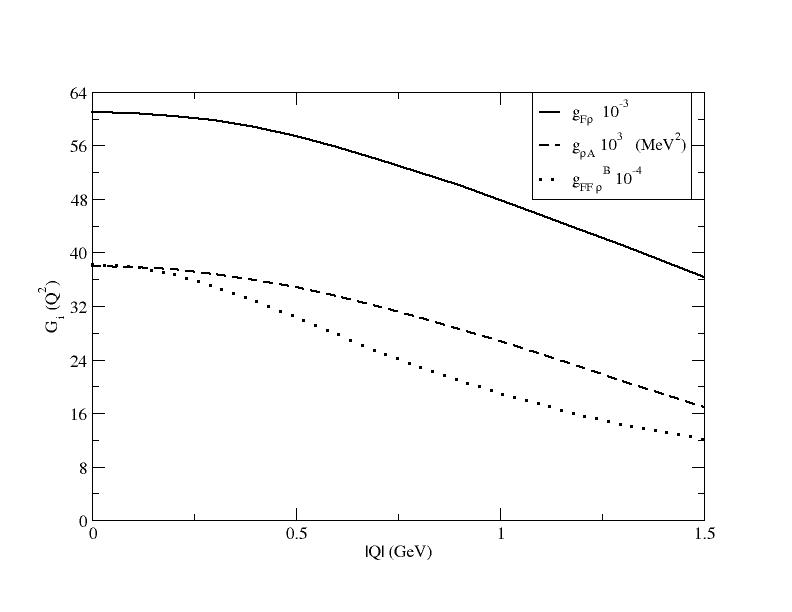}
\caption{ \label{fig:3}
\small
The form factors $G_i(Q^2)$ are the following:
$g_{\rho A}(Q^2)$  (MeV$^2$), 
$g_{F\rho}(Q^2)$ and 
$g_{FF\rho}^B(Q^2)$.
They were  presented in Eqs.  (\ref{ff-rho-B}) and (\ref{ff-F-rho-B})
and 
 (\ref{ff-FF-B}) and they
are shown for     $M^*=330$  MeV as  functions of 
spacelike $|Q| = \sqrt{|Q^2|}|$.
Dashed line is used for $g_{\rho A}$, continuous line for $g_{F\rho}$
and dotted line for 
 $g_{FF\rho}^B(Q^2)$.
 }
\end{figure}

\begin{figure}[ht!]
\centering
\includegraphics[width=140mm]{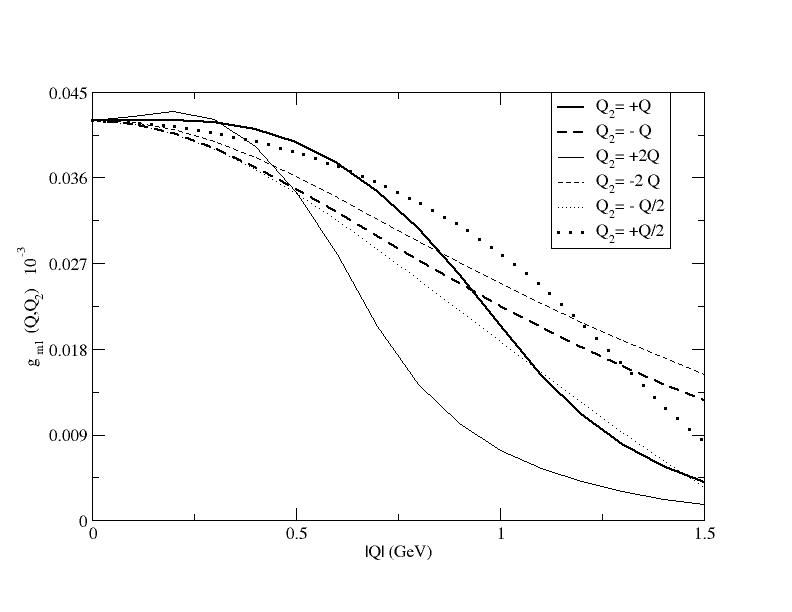}
\caption{ \label{fig:4}
\small
The form factor $g_{m1}(Q,Q_2) \times 10^{-3}$ presented in Eq. (\ref{ff-m1})
is exhibitted by considering the zero
 momentum value of the Table for $f_v=0.1$.
for $M^*=330$  MeV  as  functions of 
spacelike $|Q| = \sqrt{|Q^2|}|$.
Different particular choices for 
 spacelike momentum $Q_2$ were considered to be related to $Q$ as:
$Q_2=+Q$ means $Q_2 \cdot Q = Q^2$; 
$Q_2=-Q$ means  $Q_2 \cdot Q = -  Q^2$; 
and so one for 
$Q_2=+2 Q$;
$Q_2=- 2Q$;
$Q_2=+Q/2$;
and 
$Q_2= -Q/2$. 
}
\end{figure}

{
In Figure (\ref{fig:4})
the form factor $g_{m1}(Q,Q_2)$
is shown for $M^*=330$  MeV
and Euclidean spacelike momenta  as function 
of $|Q| = \sqrt{ |Q|^2 }$.
The regularization point shown is the one for 
$f_v= 0.1$ of the Table.
Since there are two external momenta, $Q$ and $Q_2$,
different arbitrary choices for the spacelike 
momentum $Q_2$ are exibitted.
}
However  the final values for these form factor must be multiplied by $Q$ 
due to the linear
 momentum from the strength tensor,
 and therefore it
reduces to zero for zero momentum transfer.
Firstly,  $Q_2 \cdot Q$ is considered 
to be positive
representing
a vector meson and a photon incoming to the vertex, 
and vector meson outgoing from the 
vertex.
The following different  values were adopted:
 $Q_2 = +Q$;  $Q_2=+ 2 Q$;
$Q_2=+Q/2$.
Secondly, 
$Q_2 \cdot Q < 0$ was  taken 
to represent one  incoming vector meson  to the vertex,
and a photon and the other vector meson outgoing from the
vertex.
The following values were adopted:
 $Q_2 = -Q$; $Q_2=-2 Q$;
$Q_2= -Q/2$.
There are stronger differences in the momentum dependence for
different values of $M^*$ for 
the cases $Q_2 = \pm Q/2$ and 
$Q_2 = \pm 2 Q$. 
There is a non monotonic behavior for the  $Q_2=+2Q$, and larger.

\subsection{  Pion form factor, VMD and rho-omega mixing}
\label{sec:sub-pionff}

In this section some of the above   corrections  of the $B_0$ to 
 vector meson dominance coupling
and of the $B_0$- induced vector mesons mixing are
verified on pion 
 electromagnetic  form factors.
The   pion form factor has been  parameterized
within the 
 VMD momentum assumption by the following Eq.
 \cite{PPNP-thomas}:
\begin{eqnarray}  \label{Fpi}
F_\pi (Q^2) = 1 -   Q^2  \frac{G_{vmd}}{
 Q^2 - m_\rho^2 + i m_\rho \Gamma_\rho (Q^2)}
g_{\rho \pi\pi},
\end{eqnarray}
where 
$m_\rho$, $\Gamma_\rho (Q^2)$, $g_{\rho\pi\pi}$  are
 respectively the rho mass and width 
and the rho coupling to pions.
In phenomenological models the VMD strength is 
simply described by  $G_{vmd} = g_\rho \simeq \sqrt{4 \pi \times 2}$.
The rho width can  also be considered  as momentum dependent,
by incorporating
the rho threshold, 
as \cite{PPNP-thomas}:
\begin{eqnarray}
\Gamma_\rho (Q^2) =  \Gamma_\rho 
\left( \frac{ \sqrt{Q^2 - 4 m_\pi^2}}{
 \sqrt{m_\rho^2 - 4 m_\pi^2} }
\right)^3 \left( \frac{m_\rho}{\sqrt{Q^2}}
 \right)^\lambda,
\end{eqnarray}
where $\Gamma_\rho= 146.2$  MeV \cite{PDG}
and $\lambda = 1$.

The effect of the
 weak $B_0$ induced correction to VMD momentum
dependent coupling from 
Eq. (\ref{ff})
 can be verified in the expression above.
For this, the VMD strength will be considered to be
composed by  $B_0$ independent and dependent 
components. 
{
 As discussed above the weak magnetic field contributes linearly for the 
couplings $g_{F\rho\omega}^B$ and $g_{FF\rho}^B$, therefore:}
\begin{eqnarray}  \label{gvmd-q}
 G_{vmd} (Q^2) &=&
g_{F\rho}+  
g_{FF\rho}^B (Q^2)
\frac{eB_0}{{M^*}^2}.
\end{eqnarray}

{
The rho-pion coupling constant/ form factor
 also receives a  magnetic field induced correction 
and it  can be written as:
\begin{eqnarray}
G_{\rho \pi\pi} = g_{\rho \pi\pi} + g_{F\rho\pi}^B (Q^2) \frac{eB_0}{{M^*}^2} ,
\end{eqnarray}
where $g_{F\rho\pi}^B (Q^2)$ was given in Eq. (\ref{pion-rho-B})
with therefore  the same momentum dependence of 
$g_{F\rho\omega}^B(Q^2)$.
Notwithstanding the relatively large multiplicative factor $M^*/F$ in the coupling
$g_{F\rho\pi}^B$ with respect to $g_{F\rho\omega}$ 
this coupling is somewhat suppressed.
This coupling appears strictly to one of the  vector meson (transversal) component $V_{k}^{(\mu=2)}$
and therefore can be considered to be suppressed in averaged by a constant factor
with respect to the complete pion rho coupling
$\partial^\mu (\pi^2) V_\mu^i$ for which the all the polarizations should contribute.
Besides that, in this coupling only one transversal momentum contributes in
Eq. (\ref{pion-rho-B}).
We will consider 
 the strength of the magnetic field 
correction to $g_{\rho\pi\pi}$  is reduced
by  mutiplicative  factor   $1/4$ 
basically due to the fact 
that only one polarization contributes. 
}

Besides that, a correction  due to rho omega mixing
in the vacuum  has also been 
found by considering 
the isospin breaking  up and down quark 
mass difference \cite{PPNP-thomas}.
In the present work this mixing is obtained due to
a weak external magnetic field and its isolated effect will be investigated below.
Its  contribution to be added  to $F_\pi(q)$  
{
according to phenomenological
investigations \cite{PPNP-thomas}, seemingly inspired in 
the transition $\rho^0 \leftrightarrow \gamma \leftrightarrow \omega$
 \cite{gasser-leutwyler},}   can be 
written as:
\begin{eqnarray} \label{Fpi-mix}
\Delta F_{\pi} (Q^2)  &=&
- \epsilon \;  Q^2 \frac{ g_{\rho \pi\pi} g_\omega}{
Q^2 - m_\omega^2 + i m_\omega \Gamma_\omega },
\\
\mbox{where} &&
\epsilon = \frac{ G_{\omega\rho} (Q^2) }{ m_\omega^2 - m_\rho^2
- i (m_\omega \Gamma_\omega - m_\rho \Gamma_\rho (Q^2) ) },
\end{eqnarray}
in this expression, as discussed above,  $g_\omega = g_\rho/3.5$
according to usual fits 
(considered in the numerical estimation below) 
or $g_\omega = g_\rho/3$ according to Eqs. 
(\ref{g-rho-B}) and 
(\ref{g-F-rho-B}).
The omega width is considered  to be constant and the rho-omega 
momentum dependent mixing
has a $B_0$ induced component  with a form factor 
  $g_{F\rho\omega}^B (q)$ that is proportional to $g_{FF\rho}^B$.
 It will be considered that
\begin{eqnarray} \label{Gmix+B}
  G_{\omega \rho} (Q^2) =  g_{mix} + 
 g_{F\rho\omega}^B(Q^2)
\frac{eB_0}{{M^*}^2}.
\end{eqnarray}
As reminded above the isospin breaking  mixing  strentgh 
is usually  fitted to $g_{mix} = - 4.52 \times 10^{-3}$  GeV$^2$. 
Some of the values considered below are the following \cite{PDG}:
\begin{eqnarray}  \label{values} 
m_\rho = 775 \, \mbox{MeV}, 
\;\;\;\; 
m_\omega = 783 \, \mbox{ MeV},
\;\;\;\;
g_{\rho\pi\pi} =  \sqrt{ 4 \pi \times 2.9} ,
\;\;\;\; 
\Gamma_\omega = 8.5  \, \mbox{ MeV}.
\end{eqnarray}

To show the corrections to the pion form factor induced by 
the $B_0$ dependent mixing and VMD couplings, the following differences will be
 used:
\begin{eqnarray}  \label{DF1-2}
D_1 F_\pi^2 (Q^2) &=& |F_{\pi} (Q^2)|^{2}  -| F_{\pi,B} (Q^2)|^2 ,
\\
D_2 F_\pi^2 (Q^2) &=&   | F_{\pi, mixB}  (Q^2)|^2 -  |F_{\pi} (Q^2)|^{2}  ,
\\
D_3 F_\pi^2 (Q^2) &=& 
| F_{\pi,B} (Q^2)|^2  - | F_{\pi, \rho\pi B}  (Q^2)|^2 
,
\end{eqnarray}
{where the function $|F_{\pi,mixB} (Q^2)|^2 $   contains only the 
small mixing correction 
due to the magnetic field;
 $|F_{\pi,B}(Q^2)|^2$
 includes both mixing and VMD corrections due
to the magnetic field
and  $|F_{\pi, \rho\pi B}(Q^2)|^2$ includes 
in addition to these two corrections the $B_0$ dependent
 $\rho\pi\pi$ coupling.}
In Figure (\ref{fig:5})
these functions $D_1F_\pi^2(Q^2)$,  $D_2F_\pi^2(Q^2)$ 
and $D_3 F_\pi^2 (Q^2)$ are
presented for two 
different quark effective masses,
$M^*=0.33$  GeV and $0.45$ GeV,
and for $eB_0/{M^*}^2=0.1$.
{
The overal contribution of the three magnetic field corrections
to the couplings 
is obtained by the quantity:
$D_1 F_\pi^2 (Q^2) +  D_3 F_\pi^2 (Q^2)$.
For very low timelike momenta there is a net magnetic field contribution around
$ 0.1 < Q^2 < 0.25$GeV$^2$ 
}
{
Since the correction  of the weak magnetic field is 
 of the order of $eB_0/{M^*}^2 = 0.1$
their contributions to $F_\pi^2(Q^2)$
 are at most of the order of $10^{-2}$,
and  therefore  small.
It can be seen that 
$D_1 F_\pi^2(Q^2)$ and $D_3 F_\pi^2 (Q^2)$, that 
contain
  the  VMD  and $\rho-\pi$ dependence on $B_0$,
 present   larger
deviation.
The $B_0$
 dependence of the mixing
form factor is  considerably less importante
}

\begin{figure}[ht!]
\centering
\includegraphics[width=150mm]{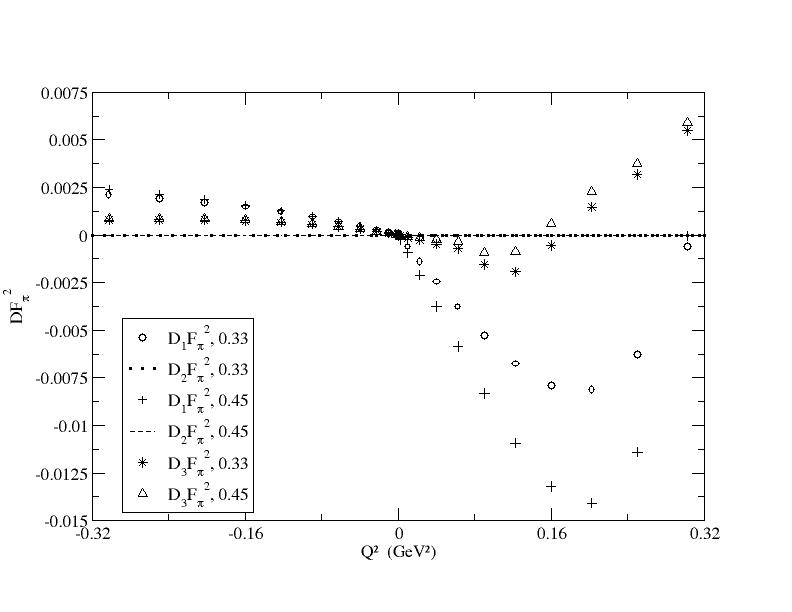}
\caption{ \label{fig:5}
\small
The difference between zero and finite weak magnetic field 
corrections to VMD and mixing strength for  
{
the fit of the  pion form factor given by Eq. (\ref{Fpi})
}
as functions of $Q^2$,  timelike ($Q^2>0$) and
spacelike ($Q^2<0$)  momenta
by means of the quantities defined in Eq. (\ref{DF1-2}).
It was considered  $(eB_0/{M^*}^2) = 0.1$ and $M^*=0.33$  GeV and $0.45$ GeV.
$m_\rho = 775$  MeV,  $m_\omega = 783$ MeV,
$g_{\rho\pi\pi} =  \sqrt{ 4 \pi \times 2.9}$,
$\Gamma_\omega = 8.5$ MeV.
}
\end{figure}

\subsection{ Charge violation potential from weak magnetic field}
\label{sec:sub-cvpot}

The charge symmetry violation  nucleon potential  has been attributed to
 isospin breaking as a source to
the light vector  mesons mixing
\cite{krein-etal-PLB,CSV-NN-1,CSV-NN-2,exp-rho-omega,hatsuda-etal,PPNP-thomas}.
Consider  a $\rho-\omega$ mixing matrix element
$ < \rho | H_{mix} | \omega >$ that has been   shown to be 
better described by a momentum dependent interaction.
Instead  of the  rho-nucleon form factor
$F_{\rho N}(Q^2)$ and omega-nucleon form factor 
$F_{\omega N} (Q^2)$  it will be considered the 
corresponding constituent quark couplings
$F_{\rho qq}(Q^2)$ and
$F_{\omega qq} (Q^2)$.
The CSV contribution in momentum space, 
for spacelike Euclidean  momenta $ Q^2 > 0$,
can be written as \cite{krein-etal-PLB,CSV-NN-1,CSV-NN-2}:
\begin{eqnarray} \label{VCSV-q}
V (Q) &=&  - 
\frac{ F_{\omega qq} (Q^2) F_{\rho qq} (Q^2) 
< \rho | H_{mix} | \omega >
}{
( Q^2 + m_\rho^2 ) (  Q^2 + m_\omega^2 ) } .
\end{eqnarray}
where 
the form factors $F_{\rho qq}(Q^2)$ and
$F_{\omega qq} (Q^2)$
were considered to be 
monopole or  quadrupole
fittings \cite{monopolar,quadrupolar,krein-etal-PLB} shown below.
In position space the CSV potential is given  by:
\begin{eqnarray}
V_{csv} (r) =  - \frac{1}{2 \pi^2 r} \int_0^{\infty}
 d \, Q \; \sin (Q r) \; Q \; 
V(Q) .
\end{eqnarray}

The effect of weak magnetic field on the mixing amplitude will be 
shown below by considering its contribution 
by means of the following quantity:
\begin{eqnarray} \label{Diff-Vcsv}
D_B V_{csv} (r) &=& 
-  (V^{(B)}_{csv} (r) - V_{csv}^{(B=0)} (r)),
\end{eqnarray}
being that   $V^{(B)}_{csv} (r)$  is calculated by adding 
the magnetic field correction to the mixing amplitude, i.e. 
\begin{eqnarray}
 \Delta_B < \rho | H_{mix} | \omega > = g_{F\rho\omega}^B(Q).
\end{eqnarray}

Two  different parameterizations  for the form factors
 $F_{\rho qq} (Q^2)$  and $F_{\omega qq} (Q^2)$ were considered: 
one with a monopole  shape
and another with a quadrupole shape 
both presented in the literature \cite{krein-etal-PLB,monopolar,quadrupolar}.
They are given respectively by:
\begin{eqnarray} \label{mono-quad-1}
F_{Mono}(Q^2) &=& \frac{ g_\rho}{ 1 + \frac{Q^2}{\Lambda_\rho^2}}
, 
\hspace{1cm}
 F_{Quad}(Q^2) =  \frac{ g_\rho }{ (1 + \frac{Q^2}{\Lambda_\rho^2})^3}
,  
\end{eqnarray}
where $g_\rho$ was given above with an analogous expressions for the $\omega$ form factors.
  The values for the constants
are
\cite{krein-etal-PLB,monopolar,quadrupolar}:
  $\Lambda_\rho = 1.4$  GeV and $\Lambda_\omega = 1.5$  GeV for 
the monopole form and 
$\Lambda_\rho \simeq \Lambda_\omega = 1.12$  GeV for 
the quadrupole form.

In Figure (\ref{fig:Vcsv-01}) 
the  magnetic field induced
contribution to the  (off-shell) CSV potential, Eq.  (\ref{Diff-Vcsv}),
$ D_B V_{csv} (r)$ is presented for
$(eB_0/{M^*}^2)=  0.1$
 by considering the two
shapes for the vector mesons form factors  (\ref{mono-quad-1})
and two quark effective masses  $M^*=0.33$ GeV and $M^*=0.45$ GeV.
For a given (fixed) effective mass
the effect of the weak magnetic field reduces mostly  to 
the multiplicative factor
$(eB_0/{M^*}^2)$
according to the expressions shown in the previous sections.
{
 Therefore for $(eB_0/{M^*}^2)  = 0.2$ or $(eB_0/{M^*}^2)= 0.3$
one can basically  multiply the values of Fig. 
(\ref{fig:Vcsv-01}) by 
2 or 3 respectively, 
with smaller correction due to the $B_0$ dependence of the 
effective mass.}
The quadrupole  form factors induces stronger suppression
of the resulting potential as it could be expected.
Therefore the difference between the two curves for $M^*=0.33$  GeV 
and $M^*=0.45$   GeV is considerably smaller for the quadrupole  form factors.
Several previous  analysis  suggested values in the  range
 $V_{csv} (0) \sim 0.1- 2.0$  MeV  for the  (total)  isospin symmetry breaking 
contribution, but a smaller  (and negative)  
strength for off-shell case, of the order of $-V_{csv} (0)
\sim 0.01-0.1$ MeV
\cite{PPNP-thomas,krein-etal-PLB,friar-etal,hatsuda-etal,CSV-NN-1}.
This  is of the same order of magnitude of the resulting $V_B(0)-V_{B=0}(0)$ 
found in Fig.(\ref{fig:Vcsv-01}) for $(eB_0/{M^*}^2)  = 0.1$.
In the case of the monopole form factors, there is 
a change of sign in the potential
around  $r \simeq 0.6$fm.
This effect had been found in different works also  for the off shell potential
with $V(r\simeq 0.75 \sim 0.9$ fm$)=0$ 
 \cite{k-dep-1,krein-etal-PLB,friar-etal,hatsuda-etal,CSV-NN-1}.

\begin{figure}[ht!]
\centering
\includegraphics[width=140mm]{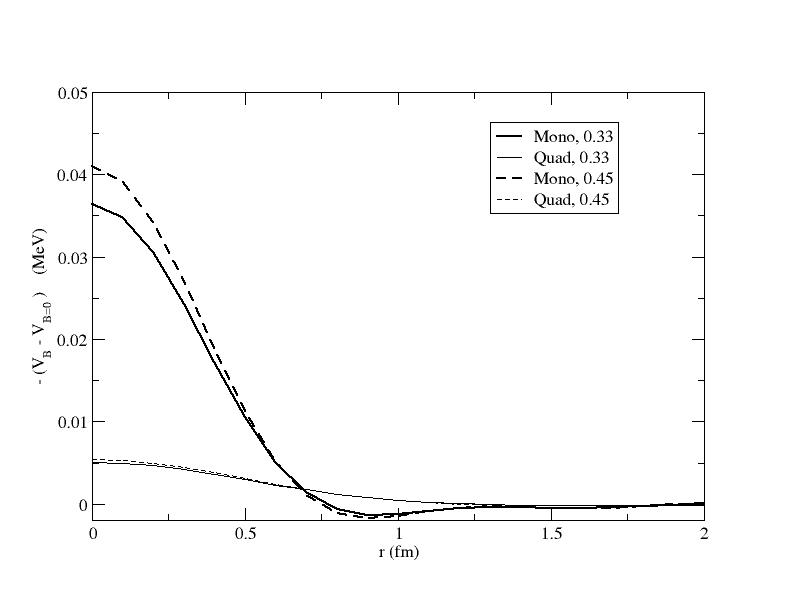}
\caption{ \label{fig:Vcsv-01}
\small
Contribution of the weak $B_0$-induced 
vector mesons mixing 
for the CSV potential from Eq. (\ref{Diff-Vcsv})
for $(eB_0/{M^*}^2) =  0.1$ and two values of the 
quark effective mass $M^*=0.33$ GeV and
  $M^*=0.45$ GeV. Both shapes for the vector mesons form factors, 
Monopole  (Mono) and Quadrupole  (Quad), 
are compared with Eq. (\ref{mono-quad-1}).
}
\end{figure}

\section{ Final remarks}

The limit of structureless vector and axial mesons considered in the present work
made possible 
  to derive simple expressions for  their leading couplings to a 
background electromagnetic field in the large quark effective mass expansion
 of the
sea quark  determinant.
The two types of phenomenological photon couplings to vector mesons (VMD),
the momentum dependent and independent ones,
and  several  other   electromagnetic  couplings were obtained.
{ From them, weak magnetic field-induced  corrections to 
usual phenomenological  coupling constants for VMD and vector
 mesons mixings were obtained.
}
A  magnetic field in the $\hat{z}$ direction
 was considered to be 
 weak with respect to an hadron mass scale such as the 
 quark effective  mass $(eB_0/{M^*}^2) < 1$.
One can choose a slightly more symmetric vector potential
such as to symmetrize the role of the transversal momentum components 
 and
vector/axial  mesons components
in Eq.  (\ref{VMD-B}).
{
Leading pion couplings to vector mesons were also found 
with their corresponding couplings to the electromagnetic field
which produce magnetic field corrections to pion-rho and pion-omega couplings.
Although these couplings are relatively larger by a factor $M^*/F$ with 
respect to VMD and mixing couplings they are suppressed due to 
reduced
polarization and transveral momenta.
}
{ 
It  can be seen that some  of the induced couplings
  are anisotropic
in momentum space due to contributions of the type  
$\partial_y V_\rho$ which corresponds to
 the plane perpendicular to the magnetic field.
All the resulting coupling constants were expressed in terms of components of   the 
quark propagator,  obtained with  DChSB, and of
the parameters of the GCM  (quark-gluon coupling constant and quark masses).
 The  non degeneracy
 of  up and down constituent  quark masses  has been investigated in several 
papers in the literature and it  
  was left outside the scope of the work to emphasize
the effect of the weak  magnetic field.
{
To render  the ultraviolet divergent  momentum integrals  finite  a 
renormalization scheme was adopted by fixing vector mesons
masses and normalization constants as renormalization conditions.
{For that,
 vector mesons and their chiral partners were assumed to 
develop the same normalization  and  
renormalization constants.}
{  Although the main aim of the work  is to derive 
photon and weak magnetic field induced
couplings between light vector and axial mesons, 
magnetic field independent couplings and free terms were 
also presented for the sake of completeness.
It turns out they provide relevant and complementary 
 information as summarized below.
The corresponding values of the VMD effective  coupling constants
in the vacuum,
$g_{\rho A}, g_{F\rho }$, 
were also calculated for the sake of completeness
and they were found to be of the order of 
magnitude of the values accepted in the literature}
being  related to 
the vector meson mass and normalization
$M_v^{(0)}$ and $g_f^{(0)}$ in Eqs. 
(\ref{grhoA-reg}) and (\ref{g-F-rho-B-reg}).
{ 
This  comparison  with usual accepted  values 
shows  it is reasonable to expect that 
$M_v^{(0)} < 0.770$  GeV and $g_f^{(0)} < 1$.
The  structureless mesons limit and isospin symmetric 
case $m_u=m_d$ should  be  responsible
for the lower values of these parameters.
}
The momentum independent   gauge non invariant
omega and rho  VMD effective 
coupling constants were  eliminated by means  of a
 field redefinition in the same way proposed  in Ref.  \cite{kroll-etal}. 
This procedure  involves
 the cancelation of a resulting 
photon effective mass from the expansion of the determinant. }
}
Light axial mesons, eventually $A_1(1260)$ and $f_1(1285)$
as  chiral partners of the vector mesons, 
{ were found to develop  similar  mixing couplings  
induced by a weak magnetic field.
 The strength of their  mixing coupling constant
was found to be slightly  larger than the 
vector mesons' one.}

}

Several  simple  relations between the resulting photon and magnetic field induced
coupling constants were found and they go 
 along the lines of the Universality hypothesis \cite{vmd}
and they present very reasonable orders of magnitude and values.
In particular, the $B_0$ correction to momentum dependent
VMD coupling constant $g_{FF\rho}^B$
 is proportional to the $B_0$ induced vector meson mixing momentum dependent
coupling constant,   $g_{F\rho\omega}^B$,
 seen in Eqs. (\ref{FFFvmd},\ref{ff-relations}) and (\ref{approx-ratios}).
These magnetic field induced corrections are UV finite.
Some simple fittings for  some of the  momentum dependent
form factors, $g_{\rho A}(Q^2)$, $g_{F\rho}(Q^2)$ 
and $g_{FF\rho}^B(Q^2)$, were presented.
The 
weak electromagnetic field expansion, as performed in this work,
accounts the 
 leading Landau orbit. 
Higher order terms of the expansion must 
provide their  complete account corresponding to 
strong $B_0$ \cite{weak-B} and then comparisons with different methods 
for the induced vector meson mixing 
 \cite{mandal-etal,rho-omega-resonance-theory} 
could be performed.

The  effect of the $B_0$ dependent  
VMD and mixing couplings  
 were investigated  in two very  different processes.
Firstly  in the 
 low momenta region of  the electromagnetic pion  
form factor for $(eB_0/{M^*}^2)= 0.1$.
{
Quite small contributions
 were found partially due to 
 to the momentum dependence of the resulting $B_0$ dependence.
Whereas the 
$B_0$-correction for the mixing yields very small contribution in 
timelike and spacelike momenta, the $B_0$-corrections
 to 
{
 VMD  and $\rho-\pi$ couplings
might yield
 sizeable contributions for very small  timelike momenta around $0.1 < Q^2 < 0.25$ GeV$^2$.}
No further
comparison with experimental data was presented because they might be obtained from
processes in which magnetic field  should not show up, in particular in $e^+e^-$ collisions.
However, the two photon coupling to vector mesons, $g_{FF\rho}$,
 might be viewed as inducing
further strength for VMD in Eq. (12) and this process might contribute in processes
like $e^+e^-$  collisions. 
A further evaluation of how this further photon coupling would
contribute by (off shell) scattering is left outside the scope of the work.
}
 Secondly, the effect of the  weak magnetic 
field on the  off-shell  charge symmetry violation 
 potential at the constituent quark level is  sizeable
and   of the order of $eB_0/{M^*}^2 $.
It is  also   considerably larger for monopole rho 
and omega form factors 
instead of quadrupole  ones, as  expected. 
Besides that, the monopole form factors also 
favor a change of sign  around $r \simeq 0.75$ fm
that had been found in several works for the
off shell CSV  potential
 \cite{CSV-NN-1,k-dep-1,krein-etal-PLB,hatsuda-etal,friar-etal}.
{
Since hadrons structure and  interactions undergo changes under magnetic fields,
  further sizeable  effects on other different components of the nucleon and 
nuclear potentials  in magnetars
 might be expected.

}

\section*{Acknowledgments}

 F.L.B. thanks   short discussions with  
F.S. Navarra,  S. Schramm and G.I. Krein. 
F.L.B. is member of
INCT-FNA,  Proc. 464898/2014-5
and  acknowledges partial support from 
CNPq-312072/2018-0 
and  
CNPq-421480/2018-1.

\end{document}